\documentclass[sigconf, screen]{acmart} %

\usepackage{xspace}
\usepackage{bbm}
\usepackage{amsmath}
\usepackage{svg}
\usepackage{xcolor,cancel}
\usepackage{mathtools}
\usepackage{dsfont}
\usepackage{wrapfig}
\usepackage{listings}

\usepackage{lipsum}
\usepackage{csquotes}

\lstdefinelanguage{InstructPseudo}{
  keywords={Guidelines, Examples},
  comment=[l]{\#},
  keywordstyle=\textbf,
  commentstyle=\textbf,
}
\usepackage{caption}
\usepackage{subcaption}
\usepackage{eucal}
\usepackage{verbatim}
\usepackage{array}

\makeatletter
\def\Hline{
  \noalign{\ifnum0=`}\fi\hrule \@height 4.\arrayrulewidth \futurelet
  \reserved@a\@xhline}
\makeatother

\copyrightyear{2025}
\acmYear{2025}
\setcopyright{cc}
\setcctype{by}
\acmConference[CHI '25]{CHI Conference on Human Factors in Computing Systems}{April 26-May 1, 2025}{Yokohama, Japan}
\acmBooktitle{CHI Conference on Human Factors in Computing Systems (CHI '25), April 26-May 1, 2025, Yokohama, Japan}\acmDOI{10.1145/3706598.3713905}
\acmISBN{979-8-4007-1394-1/2025/04}

\makeatletter
\ifcase\ACM@format@nr
\relax %
\or %
  \def\@authorfont{\large\sffamily}
  \def\@affiliationfont{\small\normalfont}
\or %
\or %
  \def\@authorfont{\LARGE\sffamily}
  \def\@affiliationfont{\large}
\or %
  \def\@authorfont{\LARGE}
  \def\@affiliationfont{\small}
\or %
  \def\@authorfont{\normalsize\normalfont}
  \def\@affiliationfont{\normalsize\normalfont}
\or %
  \def\@authorfont{\Large\normalfont}
  \def\@affiliationfont{\normalsize\normalfont}
\or %
  \def\@authorfont{\bfseries}
  \def\@affiliationfont{\mdseries}
\or %
  \def\@authorfont{\bfseries}
  \def\@affiliationfont{\mdseries}
\or %
  \def\@authorfont{\LARGE}
  \def\@affiliationfont{\large}
\or %
  \def\@authorfont{\large\sffamily}
  \def\@affiliationfont{\small\normalfont}
\fi
\makeatother
\usepackage{cleveref}

\usepackage{graphicx}
\usepackage{here}
\PassOptionsToPackage{hyphens}{url}
\usepackage{url} 
\usepackage{textcomp} %
\usepackage{multicol}
\usepackage{subcaption}
\usepackage{comment}
\usepackage{booktabs,dcolumn}
\usepackage{multirow}
\usepackage{siunitx} 
\usepackage{fancyhdr}			%
\usepackage{totpages}
\usepackage{arydshln}
\usepackage{enumitem}
\usepackage{svg}
\usepackage{rotating}
\usepackage[linesnumbered,ruled,vlined]{algorithm2e}
\usepackage[frozencache=true,cachedir=minted-cache]{minted} 
\usepackage{acmart-taps}
\aptLtoXcmd{}{

}

\def\etal{\textit{et al.}}
\def\ie{\textit{i.e.}}
\def\eg{\textit{e.g.}}

\def\systemname{InstructPipe\xspace}

\renewenvironment{quote}[1][0.04\linewidth]
  {\list{}{\leftmargin=#1\rightmargin=#1}\item\relax}{\endlist}
\newcommand{\myquote}[2]
{
\begin{quote}
\textit{``#1''} [#2]
\end{quote}
}

\usepackage{color}
\definecolor{orange}{RGB}{255,127,0}
\definecolor{darkgreen}{RGB}{0, 146, 0}
\definecolor{violet}{RGB}{148,0,211}

\definecolor{gred2}{HTML}{f6aea9}
\definecolor{gyellow2}{HTML}{fde293}
\definecolor{ggreen2}{HTML}{a8dab5}
\definecolor{gblue2}{HTML}{aecbfa}

\definecolor{gred5}{HTML}{c5221f}
\definecolor{gyellow5}{HTML}{f29900}
\definecolor{ggreen5}{HTML}{188038}
\definecolor{gblue5}{HTML}{1967d2}
\definecolor{ggrey5}{HTML}{5f6368}

\newcolumntype{L}[1]{>{\raggedright\let\newline\\\arraybackslash\hspace{0pt}}m{#1}}
\newcolumntype{C}[1]{>{\centering\let\newline\\\arraybackslash\hspace{0pt}}m{#1}}
\newcolumntype{R}[1]{>{\raggedleft\let\newline\\\arraybackslash\hspace{0pt}}m{#1}}

\newif\ifCOMMENTS
\COMMENTSfalse
\ifCOMMENTS

\newcommand{\ruofei}[1]{\textcolor{red}{{[Ruofei: #1]}}}
\newcommand{\alex}[1]{\textcolor{orange}{{[Alex: #1]}}}
\newcommand{\zheng}[1]{\textcolor{cyan}{{[Zheng: #1]}}}

\newcommand{\reviseParagraph}[1]{\begin{reviseHighlight}#1\end{reviseHighlight}}

\newcommand{\reviseSubSectionTitle}[1]{\color{blue}{#1}}

\else

\newcommand{\ruofei}[1]{}
\newcommand{\alex}[1]{}
\newcommand{\zheng}[1]{}

\newcommand{\reviseParagraph}[1]{#1}
\newcommand{\reviseSubSectionTitle}[1]{#1}
\fi

\makeatletter
\def\Hline{
  \noalign{\ifnum0=`}\fi\hrule \@height 4.\arrayrulewidth \futurelet
   \reserved@a\@xhline}
\makeatother

\begin{document}

\title[InstructPipe]{InstructPipe: Generating Visual Blocks Pipelines \\with Human Instructions and LLMs}

\author{Zhongyi Zhou$^{1,2}$, Jing Jin$^{1}$, Vrushank Phadnis$^{1}$, Xiuxiu Yuan$^{1}$, Jun Jiang$^{1}$, Xun Qian$^{1}$, Kristen Wright$^{1}$, Mark Sherwood$^{1}$, Jason Mayes$^{1}$, Jingtao Zhou$^{1}$, Yiyi Huang$^{1}$, Zheng Xu$^{1}$, Yinda Zhang$^{1}$, Johnny Lee$^{1}$, Alex Olwal$^{1}$, David Kim$^{1}$, Ram Iyengar$^{1}$, Na Li$^{1}$, Ruofei Du$^{1}$}\authornote{Corresponding author: me [at] duruofei [dot] com; Also contact: zhongyi.zhou.work [at] gmail.com}
\affiliation{
  \institution{1: Google Research, USA. 2: The University of Tokyo, Japan.}
  \country{}
}

\renewcommand{\shortauthors}{Zhou et al.}

\begin{abstract}
Visual programming has the potential of providing novice programmers with a low-code experience to build customized processing pipelines. Existing systems typically require users to build pipelines from scratch, implying that novice users are expected to set up and link appropriate nodes from a blank workspace.
In this paper, we introduce InstructPipe, an AI assistant for prototyping machine learning (ML) pipelines with text instructions. We contribute two large language model (LLM) modules and a code interpreter as part of our framework. The LLM modules generate pseudocode for a target pipeline, and the interpreter renders the pipeline in the node-graph editor for further human-AI collaboration. 
Both technical and user evaluation (N=16) shows that InstructPipe empowers users to streamline their ML pipeline workflow, reduce their learning curve, and leverage open-ended commands to spark innovative ideas.
\end{abstract}



\begin{teaserfigure}
    \centering
    \includegraphics[width=0.99\linewidth]{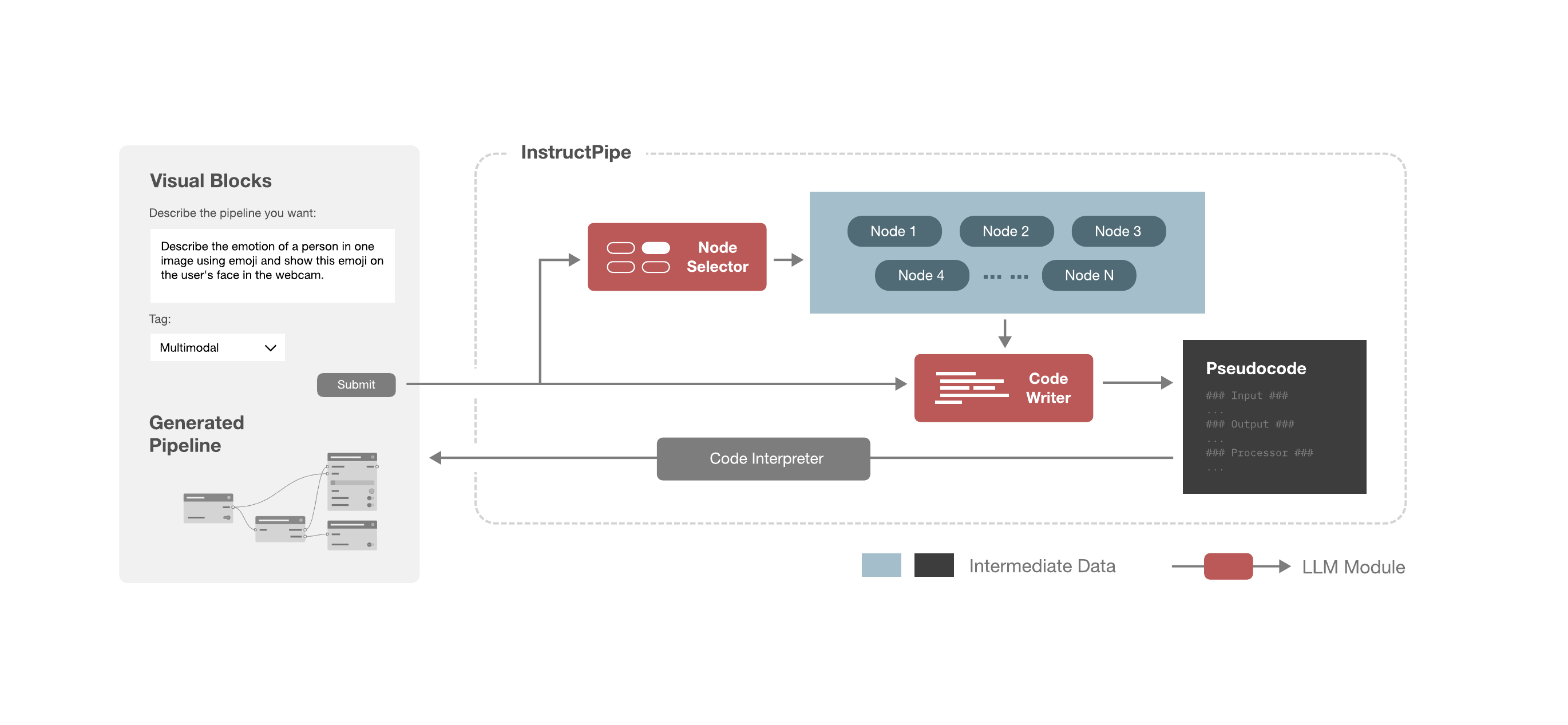}
    \vspace{0.2em}
    \caption{Workflow of InstructPipe.
    First, users describe their desired pipeline in natural language and designate it with a language, image, or multi-modal tag. InstructPipe then feeds user instructions into a node selector to identify a relevant set of nodes. Subsequently, both the instructions and the relevant nodes with their description are input into a code writer to produce pseudocode. Finally, a code interpreter parses the pseudocode, rectifies errors, and compiles a JSON-formatted pipeline, allowing users to refine and interact with it further within Visual Blocks's node-graph editor.}
    \label{fig: teaser}
\end{teaserfigure}

\settopmatter{printfolios=true}
\maketitle

\section{Introduction}
\label{sec: intro}

A \textit{visual programming} interface provides users with a node-graph editor to program through interaction with visual elements.
As opposed to writing code in a code editor, the node graph allows users to design pipelines by configuring nodes and connecting them with edges in a visual workspace.
This alternative user interface approach often accelerates experimentation and exploration in the prototyping phases of creative applications, and can make advanced technology more accessible to beginners.
Advances in machine learning (ML) further stimulate growing interest in visual programming.
Open-source ML hubs (\eg, TF-Hub~\cite{tensorflow2015-whitepaper}, PyTorch-Hub~\cite{Paszke2019Pytorch}, and Hugging Face~\cite{Wolf2019HuggingFace}) contribute large numbers of encapsulated modules that accelerate AI project development and experimentation, and such libraries provide important resources for an ML-based visual programming platform.
Recent advancements in large language models (LLMs)~\cite{Vaswani2017Attention, Brown2020Language, Anil2023PaLM} and findings on Chain-of-Thought~\cite{wei2023chainofthought} have further stimulated community-wide interest in visual programming~\cite{wu2022promptchainer, Wu2022AI, Du2023ExperiencingB, arawjo2024chainforge}, suggesting further potential in the interactive exploration of AI chains.

Despite the development of visual programming platforms in various domains, we observed that existing systems share one similar characteristic: users usually initiate a creative process in the workspace \textit{``from scratch''}.
This implies that users need to 1) select nodes, 2) ideate the pipeline structure, and finally, 3) connect nodes within \textit{a completely empty workspace}.
As was also highlighted in existing literature in programming tools~\cite{zhang2024vrcopilot, zhang2023visar}, such processes can easily overwhelm users, especially those who are unfamiliar with a particular visual programming platform.
\reviseParagraph{Providing pipeline templates may reduce on-boarding efforts~\cite{chen2024chatscratch, fogarty2001aesthetic}, but the templates inherently lack flexibility and are not easily adaptable to users’ specific needs.
}
Similar issues also arise when users write programs using text-based editors (there exist many built-in functions in a particular programming language and multiple variables in a program), but advances in LLM assistants show that such challenges can be effectively reduced. 
For example, GitHub Copilot~\cite{copilot} enables users to generate code by simply describing users' requirements in natural language.
Even though the generated code is not absolutely correct, the AI assistance usually finishes a large portion of the task, and programmers may only need to make a few edits to achieve a correct result~\cite{colab-ai, jupyter-ai}.
To this end, we raise the following question that motivates our work: \textit{How can we build visual programming assistants to accelerate the design and prototyping of ML pipelines?}

This paper introduces InstructPipe, a visual programming AI assistant that enables ML pipeline generation and design through natural language instructions. InstructPipe facilitates node connection and selection, allowing users to focus on more creative tasks like parameter tuning and interactive analysis within the visual programming workspace.
We focus our AI assistant exploration on ML-based pipelines, and therefore implemented \systemname as an extension to Visual Blocks~\cite{du2023rapsai}, a visual programming system for prototyping ML pipelines.
One major technical challenge in implementing \systemname lies in the lack of visual programming data, making it impractical to finetune a dedicated code-LLM similar to how developers build text-editor-based copilots~\cite{copilot, colab-ai, jupyter-ai}.
We addressed this issue by decomposing the generation process into three steps (\autoref{fig: teaser}).
\systemname's first LLM module scopes the potentially useful nodes, while the second LLM module generates pseudocode for a pipeline.
\systemname then parses the pseudocode and renders the pipeline in the node-graph editor to facilitate further user interaction.
Our technical evaluation suggests that \systemname reduces the necessary user interactions by $81.1\%$ when users select and connect nodes, compared to building them from scratch.
This can potentially streamline the development process, and allows users to focus on more novice-friendly interactions like parameter-tuning and human-in-the-loop verification.
Our system evaluation with 16 participants demonstrated that \systemname significantly reduced users' workload in their creative process.
Qualitative results further reveal that \systemname effectively supports novices' \textit{on-boarding} experience of visual programming systems and allows them to easily prototype concepts for various purposes.
In our experiments, we also observed new challenges caused by human cognitive characteristics, and proposed future technical directions towards open-ended AI prototyping assistants.

In summary, we contribute:
\begin{enumerate}
    \item InstructPipe, a visual programming AI assistant that enables users to generate ML pipelines from human instructions by automating node selection and connection,
    
    \item System design and technical development of InstructPipe.  The system consists of two LLM modules and a code interpreter, which generate the specification for the visual programming pipeline, compile the code, and render the pipeline in an interactive node-graph editor,
    \item Technical and user evaluations that characterize the effectiveness of~\systemname, and contribute findings that reveal new challenges for the HCI community.
\end{enumerate}

\section{Related Work}

\subsection{Visual Programming}

A computer program defines the operation of computer systems. However, \textit{``the program given to a computer for solving a problem need not be in a written format''}~\cite{Sutherland1966Line}.
This future-looking statement, dating back to the 1960s, inspired several generations of researchers to design and build visual programming systems.

Today, visual programming systems (\eg, LabView~\cite{kodosky2020labview}, Unity Graph Editor~\cite{unitygrapheditor}, PromptChainer~\cite{wu2022promptchainer}, ComfyUI~\cite{comfyui} and Visual Blocks~\cite{du2023rapsai}) typically feature a node graph editor, providing users with a visual workspace to ``write'' their program using \textit{``building blocks''}~\cite{hewett2005informing, sramek2023soundtraveller, yu2024seamez}.
Recent work further explored the application of visual programming in education~\cite{jiang2023positional, kovalkov2020inferring, chen2024chatscratch}, XR creativity support~\cite{zhang2023posevec, Zhang2020FlowMatic, ye2024prointerar}, and robotics~\cite{Datta2012RoboStudio,Huang2016Design,Huang2017Code3}.
For example, Zhang~\etal~\cite{zhang2023posevec} connected the visual programming tool to the concept of \textit{teaching by demonstration}~\cite{myers1986visual, li2005informal, zhou2022gesture}, allowing users to rapidly customize AR effects in video creation. 
FlowMatic~\cite{Zhang2020FlowMatic} extended traditional visual programming interfaces into 3D virtual environments, providing users with immersive authoring experiences.

Advancements in AI have introduced many repositories of advanced ML models~\cite{shen2024hugginggpt, huggingfacepreprocess}, and an increasing number of researchers are exploring AI chains~\cite{Wu2022AI, langchain}.
This progress has motivated HCI researchers to design and build a range of visual programming interfaces to support the AI development process~\cite{wu2022promptchainer, langflow, comfyui}.
For example, ChainForge is a web-based platform for developers to explore various LLM-related configuration and designs in a wide range of applications~\cite{arawjo2024chainforge}. 
Visual Blocks enables creation and interaction of advanced ML pipelines that can leverage state-of-the-art computer vision and computer graphics models in the browser~\cite{du2023rapsai}.

This work contributes the technical system, implementation and evaluation of a novel AI assistant that enables the use of text-based instructions in visual programming of ML pipelines.
Compared to typical workflows in which people manually build their pipelines, \systemname has the potential to accelerate ML pipeline prototyping in visual programming.

\subsection{Interactive Systems with LLMs}
\label{sec: related_llm}

The advances in LLMs bring many research directions for HCI researchers.
Researchers have started designing new LLM interfaces, to advance beyond the currently dominant chatbot interface (\textit{e.g.}, OpenAI ChatGPT, Google Gemini). %
For example, Graphologue~\cite{jiang2023graphologue} augmented LLM responses with interactive diagrams that visualize response texts in a structured format.
Sensecape~\cite{suh2023sensecape} provides users with a workspace to explore long LLM responses in a hierarchical structure.

Many HCI researchers integrated LLMs in conventional interactive systems and demonstrated that such enhanced machine intelligence can provide new user experiences~\cite{peng2023storyfier, park2022social, Liu2023Visual, wang2023enabling, feng2024canvildesignerlyadaptationllmpowered}.
This research principle is widely applied in many downstream HCI applications, including visualization~\cite{wang2023towards, shen2023towards}, explainable AI~\cite{wu2021polyjuice, wang2024human}, and social science~\cite{park2023generative, liu2023training}.
For example, Chen et al.~\cite{chen2023from} utilized LLMs to bridge low-level sensor information with high-level human requests.
Experiments showed that such connection allows users to \textit{``construct their personalized contexts [for an intelligent system] more quickly, accurately, and naturally''}.
\reviseParagraph{To interface human intention with machine operations, researchers typically utilized LLMs by following the ReAct (reasoning and acting) paradigm~\cite{yao2023react}.}
For example, Park et al.~\cite{park2023generative} simulated human behaviors in an artificial social system by leveraging LLMs as intelligent agents that perceive the environment, plan their behaviors, and act in the environment.
\reviseParagraph{Automated Visualization (AutoViz) researchers employ LLMs for data analysis and reasoning for presenting the visualization~\cite{narechania2021nl4dv, mitra2022conversationalinteraction, sah2024nl4dvllm}.
For example,  LIDA features four modules in the visualization pipeline to 1) summarize a structured dataset, 2) explore the user's goal, 3) generate code for visualization, and 4) render visualization~\cite{dibia2023lidatoolautomaticgeneration}.
ChartGPT further constructs a dedicated dataset for chart visualization, and finetunes an LLM for fully automating the data visualization pipeline~\cite{tian2024chartgpt}.
}

\reviseParagraph{\systemname extends the application of ReAct-like LLM frameworks~\cite{tian2024chartgpt, dibia2023lidatoolautomaticgeneration} to visual programming and demonstrates its effectiveness to support rapid prototyping with lower user workload.
Additionally, introducing visual programming to the ReAct framework showcases an interface solution for human-AI collaboration.
That being said, our work values partially correct AI generation, though the previous literature considers it as a complete generation failure~\cite{dibia2023lidatoolautomaticgeneration, Gupta2023Visual}. 
We leverage visual programming as a platform to integrate partially complete AI generations with human interactions, enabling even novices to intuitively collaborate with AI in their creative processes.
}

\section{InstructPipe}
\label{sect: interface}

\begin{figure*}
    \centering
    \begin{subfigure}[b]{\linewidth}
    \centering
        \includegraphics[width=0.5\linewidth]{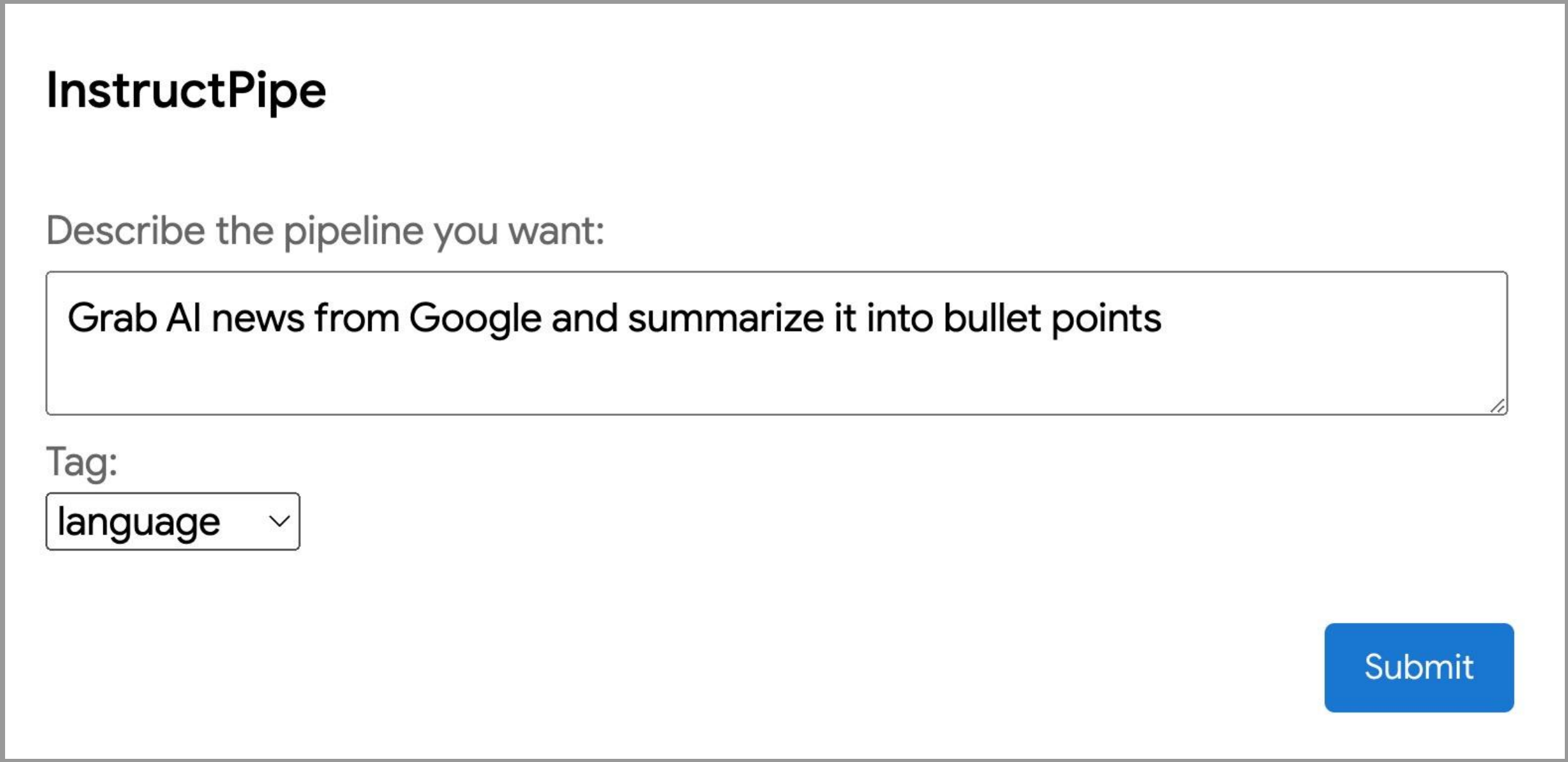}
        \caption{\systemname's instruction dialog.}
        \label{fig: dialog}
    \end{subfigure}
    \begin{subfigure}[b]{\linewidth}
    \centering
        \includegraphics[width=\linewidth]{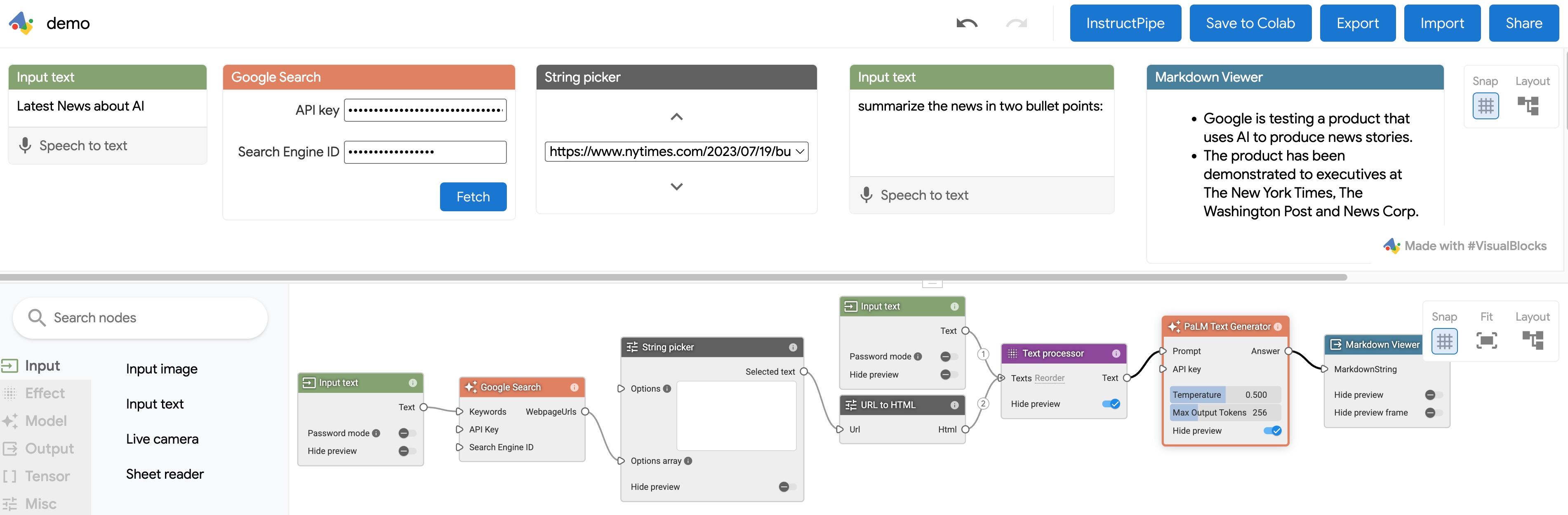}
        \caption{\systemname's visual programming interface.}
        \label{fig: interface}
    \end{subfigure}
    \caption{The user interface of \systemname. The user can first click on the ``\systemname'' button on the top-right corner of the interface in (b). A dialog will appear, and the user can input the instruction and select a category tag. \systemname then renders a pipeline on (b), in which the user can interactively explore and revise.}
    \label{fig: system}
\end{figure*}

InstructPipe is an AI assistant that enables users to generate a visual programming pipeline by simply providing text-based instructions.
We implemented \systemname on Visual Blocks~\cite{du2023rapsai, zhou2024experiencing}, a visual programming system for prototyping ML pipelines.

\subsection{User Workflow}
To generate a pipeline, users first click the ``InstructPipe'' button in the top-right corner of the interface (\autoref{fig: interface}).
The system then activates a simple dialog (\autoref{fig: dialog}) in which users provide a description and a tag for their desired pipeline.
The tag can be ``language'', ``visual'', or ``multimodal'', and helps guide the pipeline generation. 
After users click the ``Submit'' button, \systemname generates a visual pipeline in the node-graph editor.
More specifically, \systemname generates a directed acyclic graph (DAG) of a visual programming pipeline.
This implies that it uses default node parameters (e.g., the ``temperature'' or ``max\_tokens'' value of an LLM node).
Therefore, after the generation, the user needs to 1) finish the graphic structure if necessary, and 2) perform parameter tuning as well as human-in-the-loop evaluation of the pipeline quality interactively in the visual programming platform.
As we will show in our evaluation, this new human-AI collaboration approach reduces users' workload on the technical portion of the visual programming tasks (selecting and connecting nodes) and thus provides a more novice-friendly experience for technical visual programming platforms.

\subsection{Primitive Nodes}
\label{sec: nodes}

\begin{figure}[t]
    \centering
    \includegraphics[width=0.99\linewidth]{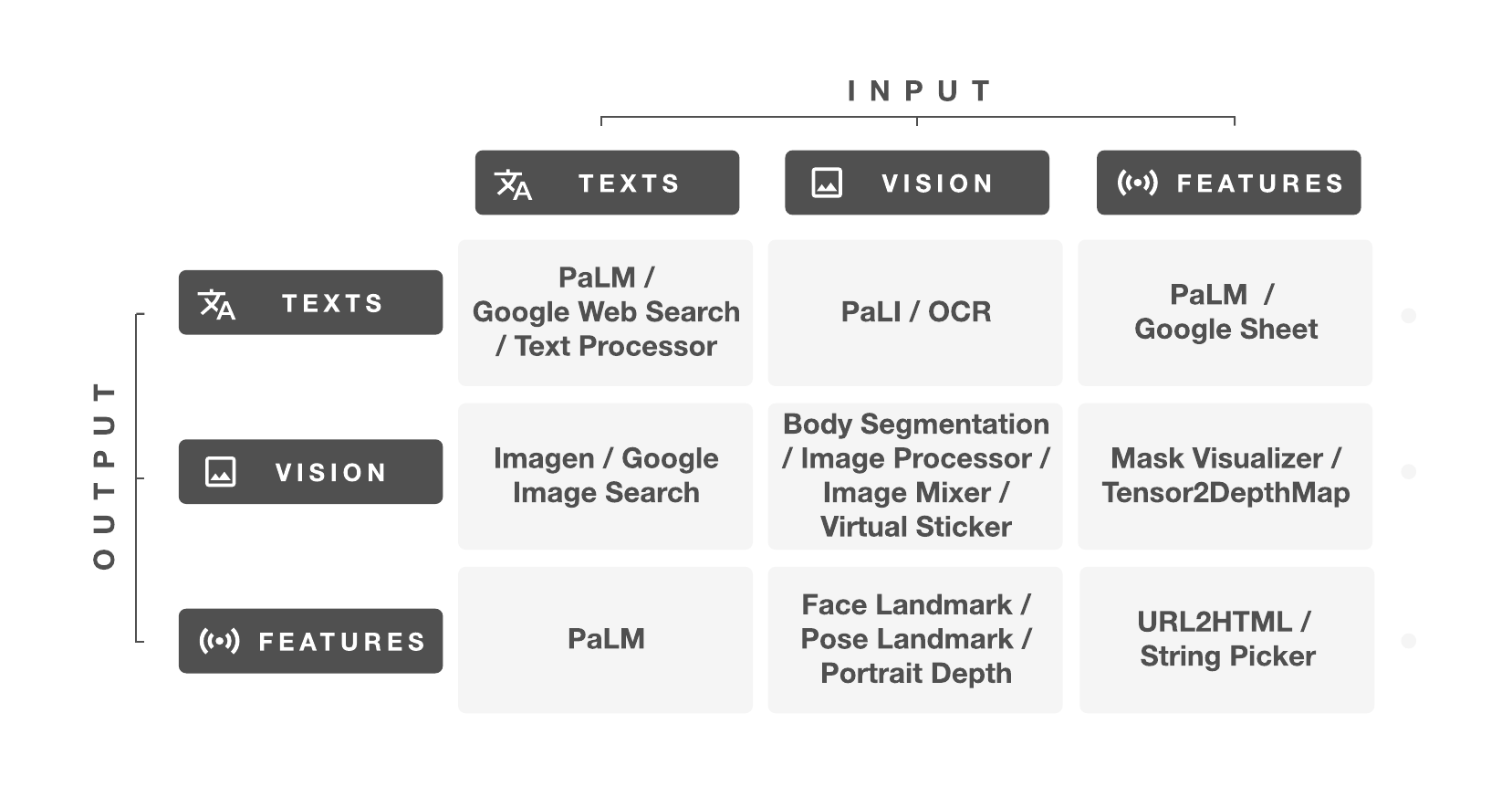}
    \centering
    \caption{
    The distribution of 20 primitive processor nodes supported by \systemname. Note that ``PaLM'' represents two nodes in \systemname, \ie, a text generation model and a chat model of PaLM~\cite{Anil2023PaLM}.   
    }
    \label{fig: primitive_nodes}
\end{figure}

InstructPipe supports 27 primitive nodes in Visual Blocks. 
We achieved this node library of \systemname by filtering out nodes without explicit definition of their functions\footnote{Note that Visual Blocks is a system that is actively being updated, and there are more nodes now.}. 
For example, `TFLite model runner' is an implicitly defined node: the user needs to input a tensorflow hub link to define its functionality.
As we mentioned previously, \systemname focuses on generating a DAG and leaves the parameter-tuning task to users.
Adding such implicit nodes without a clear definition of the functionality can easily confuse our AI assistant in the generation process, and thus, we decide to exclude these nodes in the node library of \systemname.

The 27 nodes in our library include three input nodes, four output nodes and 20 processor nodes.
The following shows an example node in each category, and we leave the full node library description in~\autoref{sec: supp-nodes}:
\begin{itemize}
    \item \textbf{``live camera'' (an input node)}: Capture video stream through your device camera
    \item \textbf{``markdown viewer'' (an output node)}: Render Markdown strings into stylized HTML.
    \item \textbf{``imagen'' (a processor node)}: Generate an image based on a text prompt.
\end{itemize}

We distributed 20 processor nodes based on the data type of its I/O edges and visualized it in \autoref{fig: primitive_nodes}.
For example. ``Google Web Search'' takes ``Texts'' information as input and output new ``Texts'', and ``OCR'' takes an image (vision-based information) as input and output ``Texts''.
``Features'' in \autoref{fig: primitive_nodes} indicates a wide range of intermediate data formats used in ML pipelines, including segmentation masks, pose landmarks, URLs and etc.
As shown in the matrix, \systemname contains a wide range of nodes that support the creation of complex ML pipelines.
Compared to related work that automates ad hoc ML inferences in specific use scenarios~\cite{Gupta2023Visual, suris2023vipergpt}, \systemname makes one more step towards the open-ended assistants with a more diverse set of primitive nodes.
Further extending our node library can effectively empower the capability of our AI assistant, which we leave as critical future work. 
In the current implementation of InstructPipe, we focus on demonstrating its capability based on our focused node library and explore what new experiences this AI assistant can bring to our users.

\section{Pipeline Generation from Instructions}
\label{sec: implementation}

\systemname leverages LLMs to generate visual programming pipelines from text instructions.
There are two prevailing approaches for LLM-customization, fine-tuning~\cite{qin2023toolllmfacilitatinglargelanguage, Liu2023Visual}, and few-shot prompting~\cite{park2023generative, Gupta2023Visual}.
Fine-tuning would require a substantial volume of annotated data, with pairs of pipelines and prompts, and it is hard to achieve for a specific visual programming platform.
Additionally, a growing list of nodes would consistently require 1) new data annotation and 2) retraining the model, making this approach less sustainable.
In comparison, few-shot prompting is a more practical approach for prototyping an interaction concept to understand the new experience it would bring to the community~\cite{Gupta2023Visual,wei2023chainofthought,yao2023react}. 
One major challenge of applying LLMs in visual programming AI assistants lies in designing efficient prompts that fit within a reasonable number of tokens. 
Even though we focus our exploration on 27 nodes, the node configuration file alone includes 8200 tokens.
Further formulating pipeline examples as in-context few-shot examples would result in a combinatorial explosion, causing an overwhelming number of tokens in the prompt. 

\begin{figure}
    \centering
    \begin{subfigure}[b]{0.99\linewidth}
        \centering
        \includegraphics[width=\linewidth]{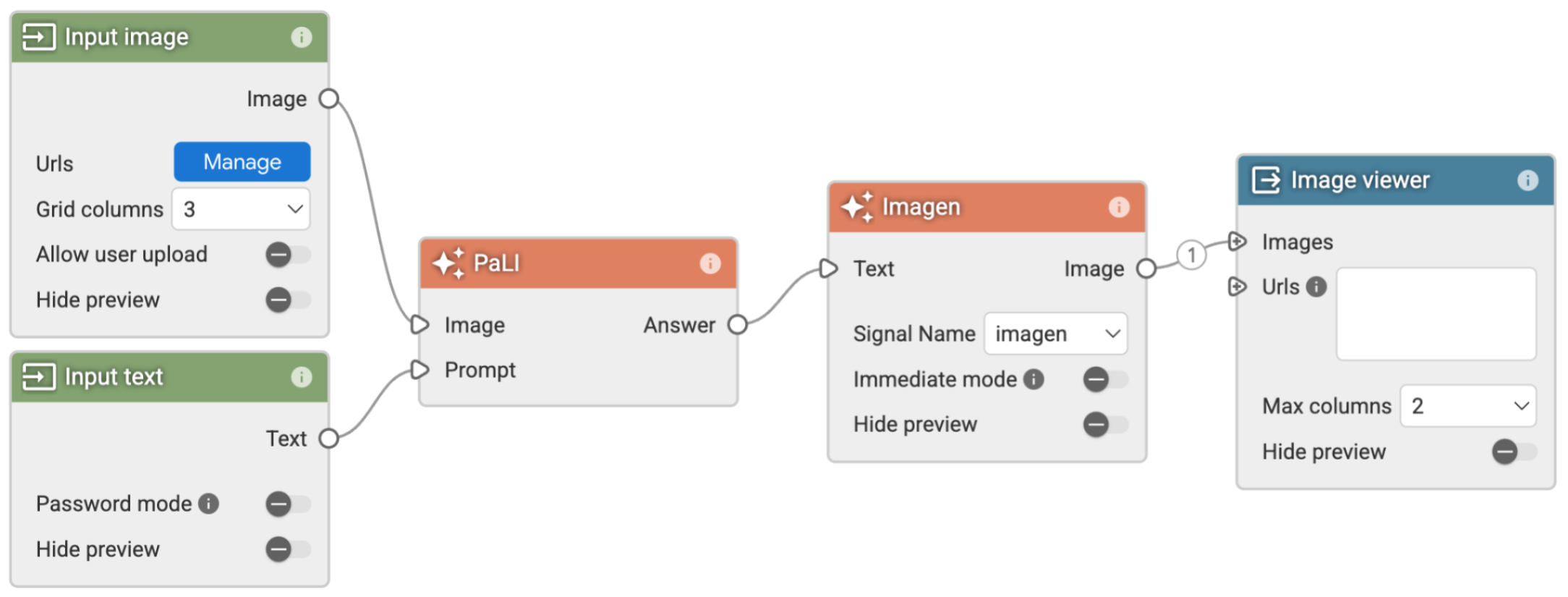}
        \caption{Pipeline.}
        \label{subfig: pipeline}
    \end{subfigure}
    \begin{subfigure}[b]{0.99\linewidth}
        \centering
        \includegraphics[width=\linewidth]{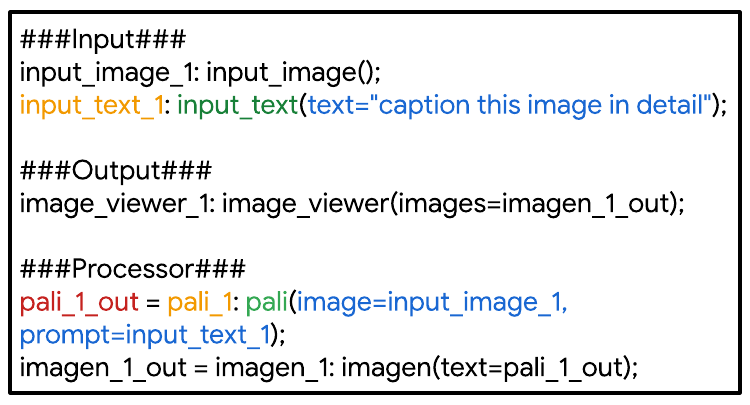}
        \caption{Pseudocode.}
        \label{subfig: code}
    \end{subfigure}
    \caption{A pair example of pipeline and pseudocode. 
    In the first line of code under ``processor'', \textcolor{gred5}{pali\_1\_out}, \textcolor{gyellow5}{pali\_1}, \textcolor{ggreen5}{pali} and \textcolor{gblue5}{image=input\_image\_1, prompt=input\_text\_1} represents \textcolor{gred5}{output variable id}, \textcolor{gyellow5}{node id}, \textcolor{ggreen5}{node type}, and \textcolor{gblue5}{node arguments}, respectively.
    }
    \label{fig: pseudo_code_example}
\end{figure}
To this end, we implement \systemname with a two-stage LLM refinement prompting strategy, followed by a pseudocode interpretation step to render a pipeline.
\autoref{fig: teaser} illustrates the high-level workflow of the \systemname implementation.
\systemname leverages two LLM modules (highlighted in red); a \textit{Node Selector} (\cref{sec: node_selection}), and a \textit{Code Writer} (\cref{sec: code_writer}).
Given a user instruction and a pipeline tag, we first devise the Node Selector to identify a list of potential nodes that would be used according to the instruction.
In the Node Selector, we prompt the LLM with a very brief description of each node, aiming to filter out unrelated nodes for a target pipeline.
The selected nodes and the original user input (the prompt and the tag) are then fed into the Code Writer, which generates pseudocode for the desired pipeline.
In Code Writer, we provide the LLM with detailed descriptions and examples of each selected node to ensure the LLM has extensive context for each candidate node.
Finally, we employ a Code Interpreter to parse the pseudocode and render a visual programming pipeline for the user to interact with.

\subsection{Pipeline Representation}
\label{sec: pipeline_representation}

The Visual Blocks system takes JSON-format data as input and renders a directed acyclic graph (DAG) in the visual programming workspace.
Therefore, the ultimate goal of \systemname is to generate the JSON file; however, directly generating the long JSON file is computationally expensive.
For example, the JSON file for rendering the pipeline in \autoref{subfig: pipeline} contains approximately 2.8k tokens.
To address this issue, we utilize the pseudocode representation of a DAG, and define this token-efficient data format as the output data format of our LLM module.
\autoref{subfig: code} shows the corresponding pseudocode representation of the pipeline in \autoref{subfig: pipeline}, and the it only contains 123 tokens.
The pseudocode representation simply stores the DAG information of a visual programming pipeline without other information such as node parameters (\eg, the ``max\_tokens'' configuration of an LLM module) and the layouts of the nodes.
This indicates that \systemname leaves the task of node parameter tuning to the user, which we believe is a more novice-friendly task, and focuses on providing technical assistance on selecting and connecting nodes.

In the following content, we provide detailed explanation on the pseudocode design and implementation.
As we mentioned above, \autoref{fig: pseudo_code_example} provides an example of a pipeline (\autoref{subfig: pipeline}) and its corresponding pseudocode (\autoref{subfig: code}).
The syntax design is inspired by TypeScript, and the overall structure is inspired by how academic papers present pseudocode~\cite{zhang2021efficient} in an algorithm block.
In \autoref{subfig: code}, we highlight the first line under the processor module (\ie, the operation of the PaLI node) in different colors, representing four different components in the programming language.
``\textcolor{gyellow5}{pali\_1}'' is the unique \textit{node ID}. 
The green symbol after the colon, \ie, ``\textcolor{ggreen5}{pali}'', specifies the \textit{node type}.
In this example, node ID  ``\textcolor{gyellow5}{$pali\_1$}'' is a ``\textcolor{ggreen5}{pali}'' node.
The arguments in brackets, \ie, ``\textcolor{gblue5}{image=input\_image\_1, prompt=input\_text\_1}'', specifies the input variables (or input edges in the graph) of this node.
``\textcolor{gred5}{pali\_1\_out}'' represents the \textit{output variable name}.
For input nodes, the output variable name is the same as the \textit{node id}, so we do not annotate the output variable with a separate name (\eg, ``\textcolor{gyellow5}{input\_image\_1}: \textcolor{ggreen5}{input\_image}()'' instead of ``\textcolor{gred5}{input\_image\_1} = \textcolor{gyellow5}{input\_image\_1}: \textcolor{ggreen5}{input\_image}()'').
Note that \systemname generates texts (\ie, the node parameter) in the ``input text'' node.
Therefore, the argument in ``\textcolor{gblue5}{text=``caption this image in detail''}'' does not indicate that the ``\textcolor{ggreen5}{input\_text}'' node accepts input edges, but accepts the node parameter input as a special case.

\begin{figure}[t]
    \centering
    \includegraphics[width=\linewidth]{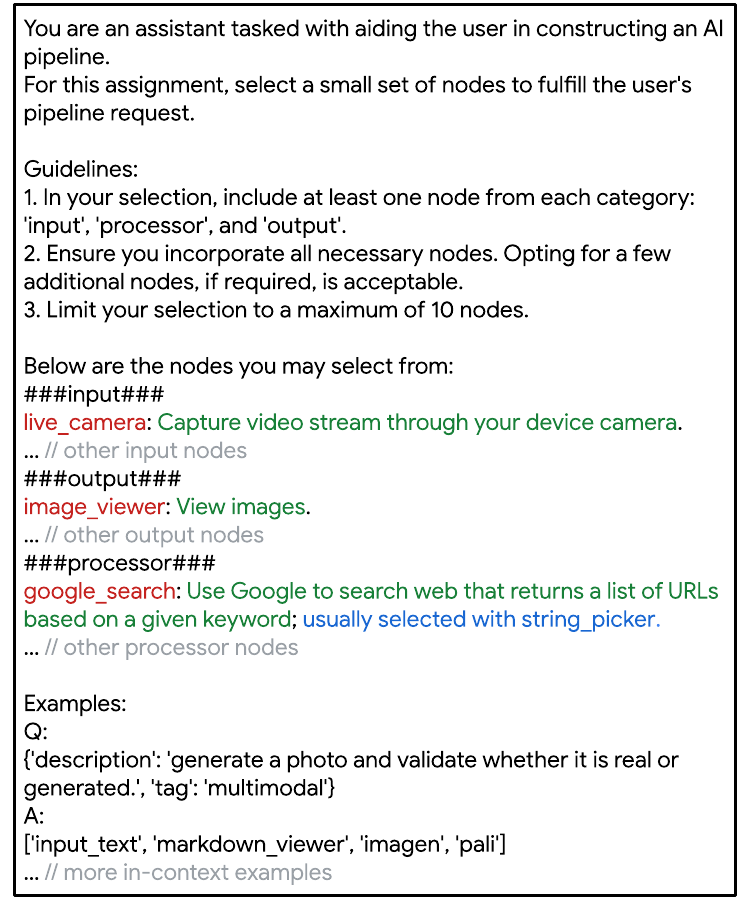}
    \captionof{figure}{The prompt structure for the Node Selection module. 
    Each node description is formated as "\{\textcolor{gred5}{node types}\}: \{\textcolor{ggreen5}{short descriptions of the nodes}\};  \{\textcolor{gblue5}{recommended node(s)}\}".
    The node recommendation is optional.
    } 
    \label{fig: node_selection}
\end{figure}%

\begin{figure}
    \centering
    \includegraphics[width=\linewidth]{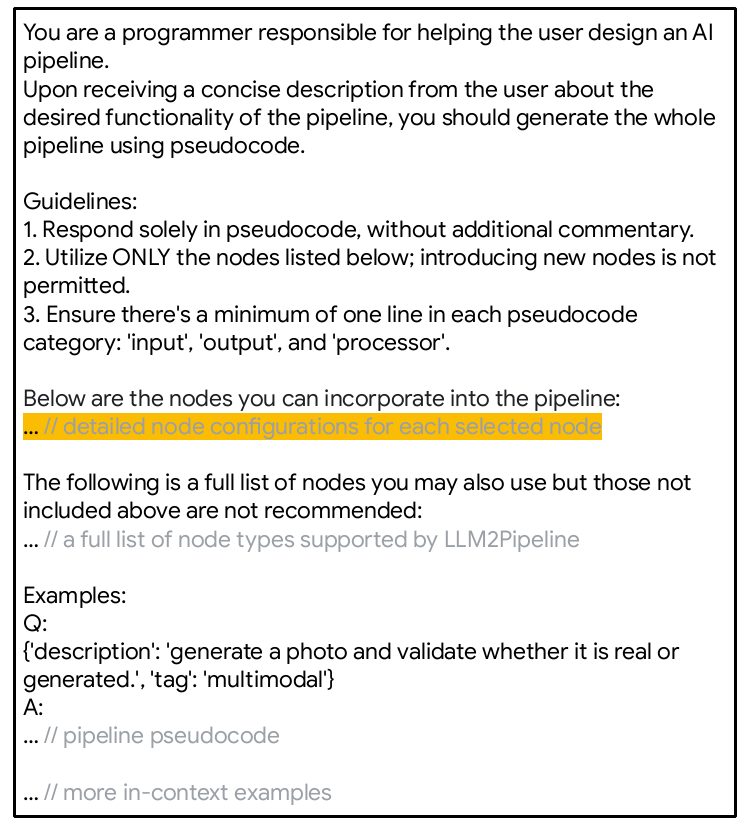}
    \captionof{figure}{The prompt structure for the Code Writer module. 
    Detailed node configurations, see the appendix for examples, are listed in the highlighted region.
    } 
    \label{fig: code_writer}
\end{figure}

\subsection{Node Selector}
\label{sec: node_selection}
Node Selector filters out unrelated nodes by providing the LLM with a short description of each node.
\autoref{fig: node_selection} shows the prompt we use in Node Selector.
Followed by a general task description and guidelines, we list all \textcolor{gred5}{node types} with \textcolor{ggreen5}{a short description} that explains the function of each.
Several nodes come with recommendation(s) when the users interact with Visual Blocks, and we also include such \textcolor{gblue5}{node recommendations} in the prompt.
The main intuition of this prompt design is based on how existing open-source libraries (\eg, Numpy~\cite{harris2020array}) present a high-level overview of all functions\footnote{See an example in the following link: \url{https://numpy.org/doc/1.25/reference/routines.array-manipulation.html}}.
The documentation typically presents a list of supported functions (in each category), followed by a short description so that developers can quickly find their desired functions.
At the end of the prompt, we provide a list of Q\&As as few-shot examples to support the LLM to learn and adapt to the context of the task.

\subsection{Code Writer}
\label{sec: code_writer}

With a pool of selected nodes, the Code Writer module can write pipeline rendering pseudocode.
\autoref{fig: code_writer} shows the structure of the prompt utilized in this LLM module.
Similar to \cref{sec: node_selection}, the prompt starts with a general introduction and several guidelines.
The major difference in the prompt design in this stage lies in the granularity of each node introduction.
We provide a detailed configuration for each selected node with additional information, including 1) input data types, 2) output data types, and 3) an example, represented in pseudocode, for how this node connects to other nodes.
We include a detailed explanation of the full node configuration in~\autoref{appendix: code_writer}.
Similar to the previous LLM module (\cref{sec: node_selection}), the prompt design here is also inspired by the documentation of existing software libraries. 
Specifically, we gain inspiration from low-level function-specific documentation\footnote{See an example in numpy.shape: \url{https://numpy.org/doc/1.25/reference/generated/numpy.shape.html\#numpy-shape}}, which typically includes 1) a detailed description, 2) data types in the input/output, followed by 3) one or more examples of a few lines of code for how developers can use the function.

The prompt also includes a Q\&A list as few-shot examples.
However, providing few-shot examples in this stage is non-trivial.
The reason lies in the dynamics of the node selection pool. 
A combination of all the nodes causes many possible options, and it is impossible to design a dedicated list of few-shot examples in each possible case.
Therefore, we only created an example list for each pipeline tag (\ie, ``language'', ``visual'', and ``multimodal'') and intended to utilize these few-shot pipelines to teach LLMs example use cases in each category.
This implies that in-context pipelines may include nodes that were not selected for the prompt.
This can potentially lead to LLM hallucinations~\cite{huang2023surveyhallucinationlargelanguage}, \ie, utilizing the nodes that do not exist in our node library.
We mitigated this issue by adding specific prompts that explicitly show a list of supported nodes (\ie, the contents start with ``the following is a full list of ...'' in \autoref{fig: code_writer}).
However, LLM hallucination is a community-wide challenge, and we also find that our approach cannot eliminate this issue in visual programming.
Therefore, \systemname conducts a sanity check for the Code Writer outputs and directly disposes of the line of pseudocode with such hallucinated nodes.
This can ensure that the generated code is in a valid data format for rendering the pipeline in Visual Blocks.

\subsection{Code Interpreter}
\label{sec: code-interpreter}

After our LLM modules generate the pseudocode, InstructPipe employs a code interpreter to parse the generated pseudocode and compile a JSON-formatted pipeline with an automatic layout. 
Since we incorporated standard approaches to achieving such conversion from the pseudocode to the JSON file, which we do not intend to claim as our main contributions, we briefly summarize our implementation into the following three steps for simplicity~\reviseParagraph{and elaborate low-level implementation details at Appendix~\ref{appendix: code_interpreter}}:

\begin{enumerate}
    \item \textbf{Lexical Analysis}: \systemname first tokenizes each line of the pseudocode into \textcolor{gred5}{output variable id}, \textcolor{gyellow5}{node id}, \textcolor{ggreen5}{node type}, and \textcolor{gblue5}{node arguments} (\cref{sec: pipeline_representation}).
    \item \textbf{Graph Generation with Default Node Parameters}: We generated a DAG based on the tokenized results and applied predefined default node parameters in each generated node.
  
    For example, by default, the temperature and the max output tokens for the PaLM node are set to 0.5 and 256, respectively.
    If users are not satisfied with the default values, they can interactively adjust the parameters in the node-graph editor.
    \item \textbf{Layout Optimation}: 
    When pseudocode is converted into a JSON file, default node parameters will cause sub-optimal visual effects (\autoref{fig: smartlayout_before}).
    \systemname conducts a layout optimization process using the breadth-first search (BFS) algorithm, which re-arranges the layout for better presentation of the pipeline (\autoref{fig: smartlayout_after}).

\end{enumerate}

\section{Technical Evaluation}
\label{sec: tech_evaluation}

\systemname contributes a framework for generating specifications for visual programming pipelines based on text prompts from users. To characterize the system's performance, we designed a technical evaluation to assess the accuracy of the generated graphs for a variety of prompts.

\subsection{Data Collection}
\label{sec: data_collection}
To compute the accuracy of our generated pipelines, we need to collect a corpus with pairs of instructions and their corresponding ground-truth pipelines.
Therefore, we organized a two-day hybrid workshop with 23 participants, aiming to collect real pipelines that Visual Blocks users would build for their creative usage.
The 23 participants (F: 6; M: 17) are composed of five software engineers, four research scientists, four students, three designers, two project managers, and two engineering managers. 
Six attendees claimed that they had prior experience in using Visual Blocks.
As this was a data collection study rather than a user study, where each participant here served as a data creator and annotator, we did not restrict participation to individuals who self-identified as novices.
The workshop began with a 15-minute Visual Blocks tutorial walking the participants through the nodes and the pipeline-building process.
After the tutorial, attendees created pipelines independently. 
Once they finished creating the pipelines, participants were required to caption their pipelines and upload them.
We utilized this corpus of data pairs (caption/pipeline) as the data set for the technical evaluation.

The workshop was an open-ended creation process in which participants were free to use any node available in Visual Blocks with more than the 27 nodes covered by \systemname.
\textit{The \systemname feature was not available in the workshop}.
After the workshop, we post-processed our collected data and achieved 48 pipelines (23 language pipelines, seven visual pipelines and 18 multi-modal pipelines) for our technical evaluations. 
The post-processing procedure details are presented in 
Appendix \ref{sec: tech_postprocess}.

\subsection{Metric: The Number of User Interactions}
\label{sec: metric}
To quantify the efficacy of \systemname based on our goal of accelerating and streamlining pipeline creation, we defined the metric \textit{Number of User Interactions} as follows:

\begin{displayquote}
\textit{The Number of User Interactions is defined as the \textbf{minimal} number of user interactions needed to \textbf{complete} the pipeline from a generated pipeline.}
\end{displayquote}

This definition is mainly inspired by Graph Edit Distance (GED) in graph theory~\cite{gao2010survey}.
Note that there are countless ways to modify a generated pipeline toward a complete pipeline in practice.
Nevertheless, the \textbf{minimal} number of user interactions is deterministic, and this is an objective metric that can fairly estimate the amount of effort users need to spend to achieve their goal.
A pipeline is considered \textbf{complete} when it satisfies the given instruction.
We calculate the number of interactions across two types of events: 1) adding/deleting a node, and 2) adding/deleting an edge between nodes.
In the technical evaluation, we report the average \textit{ratio} of user interactions required to complete a pipeline ``from our generated pipeline'' compared to ``from scratch'' as our target metric.
For example, if it takes 3 interactions to complete a pipeline from our generated results and takes 10 interactions to complete from scratch, then the ratio of interactions is $30\%$.
Appendix~\ref{appendix:metric} contains further discussion of this metric.

\begin{table}[t]
\caption{The ratio of human interactions in the technical evaluation. Results are reported as mean $\pm$ standard deviation. }
\label{tab: tech_results}
\vspace{0.3em}

\begin{tabular}{cccc}
\toprule
\textbf{Overall}  & Language &  Visual &  Multimodal      
\\ \midrule      
$\mathbf{18.9\pm 20.3}\textbf{\%} $      & $17.4\pm 20.6 \%$        & $17.6\pm23.7\%$  & $20.8\pm 16.0\%$          \\ 
\bottomrule  
\end{tabular}

\end{table}

\subsection{Experiment Setups and Results}
\label{sec: tech_results}
We ran our generation algorithm on the pipeline captions six times (three times for each caption $\times$ two captions for each pipeline), and computed an averaged performance among the six trials for each pipeline.

Table~\ref{tab: tech_results} summarizes the results of the technical evaluation.
Compared to building a pipeline from scratch, \systemname allows the user to complete a pipeline with \textbf{18.9\%} of the user interactions, demonstrating the potential of \systemname to require more than 5X fewer interactions.
\textbf{Seven} generated pipelines directly satisfied with instructions without user interactions in all six trials, and \textbf{38} generated pipelines completed at least once in any of the six trials.

\section{User Evaluation}
\label{sect: system_eval}

While the technical evaluation demonstrates the accuracy of \systemname among various real pipelines created by participants, it is still unclear what is the actual user experience when real users go through the entire system workflow.
Therefore, we conducted an in-person user study of \systemname with another group of participants, aiming to provide more insights into our system performance as well as explore new user experiences brought by \systemname.
The study recruitment was in accordance with the ethics board of Google.
We obtained participant consent before the study began.

\subsection{Study Design}
\begin{figure}
    \includegraphics[width=0.5\linewidth]{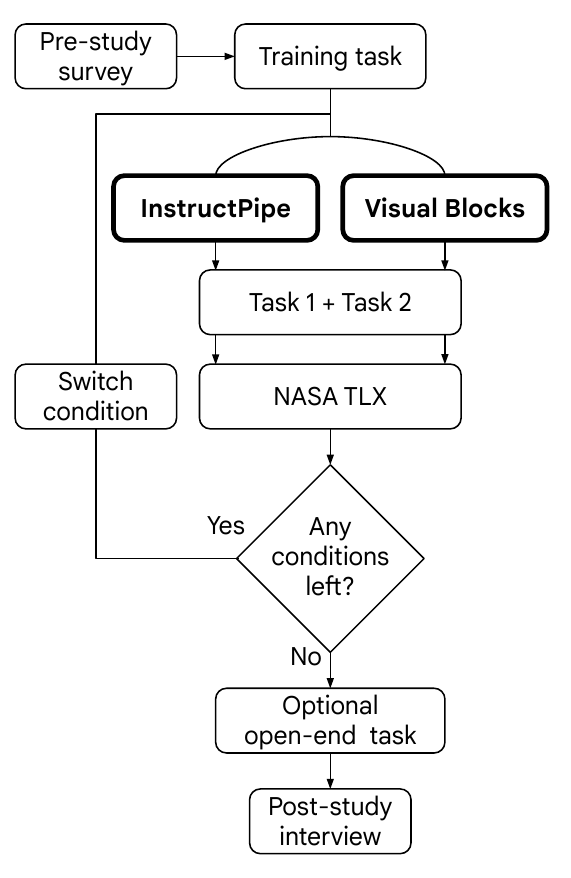}
    \caption{A flow diagram of the user study. After a training session, participants completed the two tasks in each condition in the sequence determined by the counterbalancing protocol.
}
    \label{fig:user_evaluation_flow}
\end{figure}

In the user evaluation, we aimed to investigate how the interface condition (with \systemname and without \systemname; the independent variable) affects the user experience and behaviors (the dependent variable).
We will refer to these two interface conditions as ``\systemname'' and ``Visual Blocks'' in the following content.
\autoref{fig:user_evaluation_flow} visualizes the complete study flow. 
In each condition, participants completed the two pipelines with counterbalance (referred to as Task 1 and Task 2 in \autoref{fig:user_evaluation_flow}).

We carefully designed the experiment to create a fair study that could be completed with reasonable effort. 
In the following content, we elaborate on how we made two important decisions related to the study's rigor:

\subsubsection{\textbf{\reviseSubSectionTitle{Two controlled pipelines with full counterbalancing}}}
Our user evaluation focuses on two controlled pipelines \reviseParagraph{with full counterbalancing.}
While we acknowledge that more pipelines (\eg, four, six, or more) could enhance generalizability, such designs would also inevitably increase the size of the required user groups\reviseParagraph{, even without fully counterbalancing.
For example, fully counterbalancing four controlled pipelines requires $12 \times$ more participants. 
Partially counterbalancing with four pipelines using the Latin Square design still requires us to double the number of participants.
Additionally, novice participants are likely to progressively gain experiences within the study, and such learning effects will weaken the design of partial counterbalancing.
We believe that two pipelines with full counterbalancing are a reasonable experiment setup in this work, and future work could consider extending and scaling up these experiments. 
}

\subsubsection{\textbf{Pipeline selection}}
\reviseParagraph{Given the fixed number of pipelines we can evaluate with users and the potential bias introduced by few-shot prompts~\cite{calibrate2021zhao}, it is important how we select the two pipelines for user study.}
There are two critical factors that we considered: representativeness and diversity. 
Representativeness implies that the selected pipelines should represent the average performance of \systemname.
Diversity suggests that the selected pipelines should provide various experiences to simulate the actual use scenarios in which the performance of \systemname may vary.
Following this guideline, we selected four candidates, and the final decision was made after a pilot study with one participant to test the level of pipeline difficulty.
The two resulting pipelines are composed of eight nodes with seven edges, and six nodes with six edges, respectively.
Using the instructions from two authors, the averaged ratio of human interactions in these two pipelines are $27.8\%$ and $5.2\%$, respectively.
See \autoref{appendix: user-study-pipeline} for more detail on the pipelines.

\subsection{Procedure}
\label{sec: procedure}

Each study session takes 55 - 65 minutes in total.
The study started with 10-15 minutes of hands-on training for both conditions. 
The training included 1) all the Visual Blocks interactions needed to complete the subsequent steps of the experiment, and 2) all the nodes that participants will need to use for pipeline creation in the main session.
Participants were also encouraged to experiment with building a pipeline independently, and to ask questions.

After the training, participants progressed to a formal study session where they were asked to build and complete pipelines under the given conditions.
We verbally described the pipelines to participants as below, and participants could not see our scripts:
\begin{itemize}
    \item \textbf{Text-based pipeline}: get the latest news about New York using Google Search and compile a high-level summary of one of the results.
    \item \textbf{Real-time multimodal pipeline}: create a virtual sunglasses try-on experience using your web camera.
\end{itemize}
A pipeline is considered complete when the aforementioned functions run in the user's visual programming workspace.
For example, we consider the ``real-time multi-model pipeline'' as complete when the pipeline registers the sunglasses on the user's face, with real-time tracking and following of the head movement.

During the task, participants were allowed to consult with us for technical help. 
If participants were unable to make progress, we provided hints. 
We provided many more hints in the baseline condition, and we made this decision to ensure every novice-level participant can finish their tasks within a reasonable amount of time.
Appendix~\ref{appendix: user_study_assistant} contains more details and discussions of the assistance we provided in the study.
As an optional extension to the study, eight participants were offered an open-ended pipeline creation, where participants prototyped their own ideas using \systemname. 
This optional section was offered based on the progress of the participant in the previous sections, and time constraints so that the study duration was controlled within the time we guaranteed in our recruitment process.

After conducting all pipeline-condition combinations, participants answered open-ended questions in a semi-structured interview. The interview script is available in the appendix \ref{sec:supp-interview-script}. 
Participants provided their general impression of each condition, listed pros and cons, identified potential future use cases, and critiqued the user interface for future improvements. 
\reviseParagraph{We transcribed the interviews and conducted the open coding analysis on the qualitative data~\cite{strauss1987qualitative, strauss1990basics}.
More specifically, we categorized the quotes based on our observations and then refined the code for presentation.}

\subsection{Participants}
\label{sec: }

We recruited 16 participants from our internal participant pool, which is specifically designed for UX research within our institution. Importantly, none of the participants was involved in our project, and the authors in charge of the study did not personally know any of the participants.
We screened participants on their self-reported programming experience and machine learning skills.
All of the 16 selected participants rated their ``Programming Experience'' and ``'Machine Learning Skill' as ``Intermediate'' or below (See \autoref{fig:ppt_demo} for a full breakdown).
We intentionally recruited novice users, as we envision them as intended users of InstructPipe.

\subsection{Metrics}
In addition to the qualitative data from the interview, we measured the following quantitative data. 

\subsubsection{Task Completion Time}
Back-end logs were used to collect timestamps for starting and ending events. Then, the completion time for each condition was calculated per task for each participant. 

\subsubsection{The Number of User Interactions}
We used the number of user interactions (introduced in \cref{sec: metric}) to measure the user's objective workload.
Unlike the results in~\cref{sec: tech_results}, we report an absolute value here because all the pipelines are controlled in the system evaluation.

\subsubsection{Perceived Workload}
The raw task load index (Raw-TLX) questionnaire was used to measure participant's perceived workload~\cite{hart1988development}. 
This questionnaire was a subset of the NASA-TLX (part I).
Participants filled out the questionnaire after each condition (\systemname or Visual Blocks). 

\subsection{Results}
\label{sec: sys_result}
\begin{figure*}
    \centering
    \includegraphics[width=\linewidth]{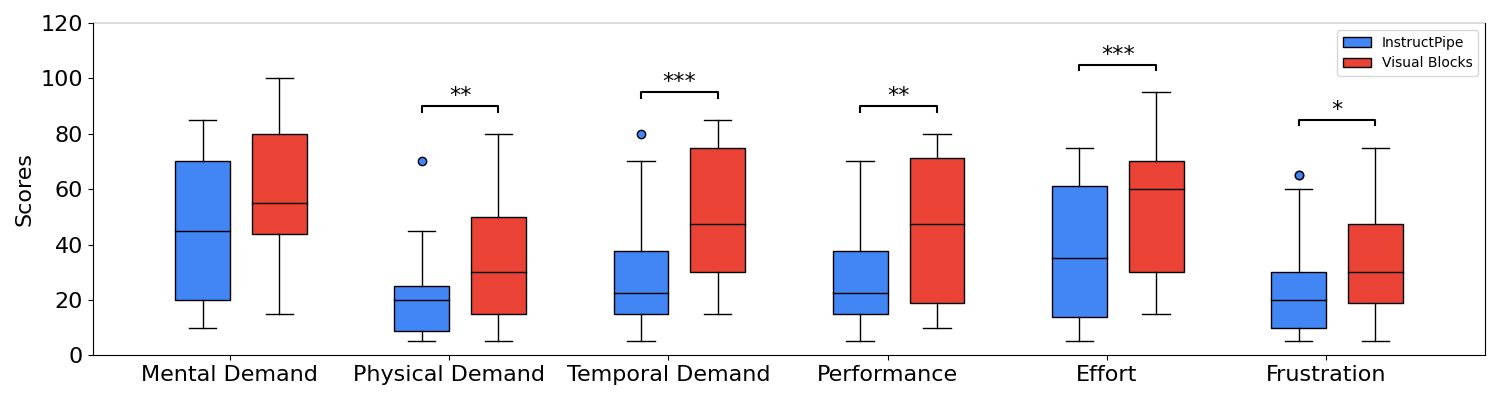}
    \caption{Raw-TLX results.
    The statistic significance is annotated with $^*$, $^{**}$, or $^{***}$ (representing $p$<.05, $p$<.01, and $p$<.001, respectively).}
    \label{fig: tlx}
\end{figure*}

\begin{table}[t]
\caption{Task completion time and the number of human interactions in the user study (N=16). 
We use $***$ to denote $p < .001$.
}
\label{tab: sys_results}
\vspace{0.3em}
\begin{tabular}{ccccccc}
\toprule
\multirow{2}{*}{System} & \multicolumn{3}{c}{\textbf{Time} (secs)}                     & \multicolumn{3}{c}{\textbf{\# Interactions}}           
\\  
                        & Median & IQR    & p                                 & Median & IQR  & p                                 \\ \midrule 
InstructPipe            & \textbf{203.5}  & \textbf{156.25} & \multirow{2}{*}{***} & \textbf{5.0}   & \textbf{4.25} & \multirow{2}{*}{***} \\
Visual Blocks           & 304.5  & 124.25 &                                   & 16.0   & 6.0  &                                   \\ 
\bottomrule 
\end{tabular}
\end{table}

\subsubsection{\textbf{\systemname Reduces Users' Workload}}
\label{sec: workload}
Table~\ref{tab: sys_results} shows the results of two objective metrics measured in the study.
The Wilcoxon signed ranks test found significant differences on both scales ($p< .001$).

\autoref{fig: tlx} further visualizes the results of users' perceived workload in six sub-scales.
The Wilcoxon signed ranks test revealed significant differences on five sub-scales, all except ``Mental Demand'' (see \cref{sec: mental_workload} for more explanations and discussion).
Furthermore, the test indicates that all participants unanimously considered that \systemname provides lower or equal workload on the subscales of ``Physical Demand'', ``Temporal Demand'', ``Performance'' and ``Effort'' ($W=0$).
These quantitative results, with both objective and subjective metrics, demonstrate the potential for \systemname to dramatically reduce users' workload during the pipeline creation process.

Users' qualitative feedback is also aligned with our quantitative results. 
Participants complimented that \systemname is \textit{``helpful''} [P16] and \textit{``obviously easier (to use) than [Visual Blocks]''} [P1].
P11 and P6 further elaborated how \systemname enhances the user experience when the user builds a visual programming pipeline:
\myquote{I feel like I can talk in natural language, and it (InstructPipe) can write the code for me.}{P11}

\subsubsection{\textbf{On-boarding Support of Visual Programming}}
\label{sec: onboarding}

P1, P5, and P9 explicitly mentioned that there is a \textit{``learning curve''} in visual programming systems, which validates our statements and motivation in~\cref{sec: intro}.
\myquote{There is a learning curve to it (using the visual programming system) for sure, because you have to, like, read each node carefully.}{P1}

P1's comment matches our observation of participants' behaviors during the study. 
In the Visual Blocks condition, we observed that people were more easily stuck in their creative purposes, which required our support.
Typical support included 1) guiding participants if they went too far away from the correct pipeline, and 2) reminding them of an important node for the pipeline, although we introduced all the necessary nodes in our training session.

To this end, participants commented that \systemname is a good onboarding tool in visual programming systems, especially for non-experts, to get familiarized with the system by having a ready solution.
\myquote{[InstructPipe] lets you know these nodes exist [when the pipeline appears after the instruction]. It’s like a super speedy tutorial.}{P7}
\myquote{If you don’t have experience in visual programming, you will appreciate [InstructPipe] much more ... With [InstructPipe], the structure is there, and I feel less worried about making mistakes. It’s, like, giving you examples. It’s easier than starting from scratch.}{P5}

Anecdotally, three participants asked for \systemname during the Visual Blocks condition.

\subsubsection{\textbf{Integration into the Existing Workflow}}
\label{sec: integration}
\systemname is a feature available in Visual Blocks. 
In the interviews, participants particularly expressed their strong appreciation of this design as an AI assistant that enhances, instead of completely replacing, the existing user workflow:%

\myquote{[The pipeline generated by~[InstructPipe] could be pretty close to what I want ... Or maybe sometimes not, but that's okay. I got most of the blocks there, and then it's up to me to figure out how to connect them.}{P6}

While most participants, like P6, appreciated the integration of the AI assistant into the standard visual programming workflow, P15 expressed a concern about this approach. In visual programming, users typically rely on visual thinking to construct pipelines, but the new prompt-based method introduces a shift toward text-based reasoning. This blend of cognitive processes could potentially increase users' mental workload:
\myquote{
[the participant is talking about s/he wants to fix an unsuccessful generation by changing the prompt instead of performing visual programming here] ... because I just spent so much time figuring out what the prompt should be. That's kind of like \textbf{already where my brain was} and I knew that something was wrong there (the prompt), but I would \textbf{have to switch over to the other mode} (visual programming) of figuring out what was wrong in the pipeline ... [this is very overwhelming]}{P15}

\subsubsection{\textbf{Use Scenarios: Accessible ML Prototyping and Education}}
\label{sec: use_scenario}
In the open-ended session, we observed that participants could efficiently utilize \systemname to prototype a pipeline for various daily life or business purposes.
For example, 
P14 tried InstructPipe with \textit{``summarize real estate price increase in San Diego California over 2023''}. 
Compared to using LLM chatbots, \systemname helps the user quickly build a more explainable pipeline in which the user can track (or modify) the information resources.
P4 prototyped an interactive VQA app by ``\textit{Describe the product in the camera}''.
P13 further shared his thoughts on how this rapid and accessible prototyping experience can support future business:

\myquote{It (\systemname) is going to facilitate prototype building for PMs (Product Managers) ... I have lots of ideas, but my challenge is how to translate an idea into the technical world and see a prototype. I think that this app expedites me in that process a lot.}{P13}

Another emerging theme was regarding educating kids on programming: \myquote{With [InstructPipe], I don't need to teach them (kids) to code for them to build something
... Some kids like to code, some kids like to create stuff but don't want to be bored with learning the syntax of coding ... Using [InstructPipe], I can see kids can build, like, customized chat-bots or interactive Wikipedia.}{P13}

\subsubsection{\textbf{Limitations and Future Directions}}
\label{sec: feature_request}

Across the study sessions, we consistently observed a specific user behavior pattern: participants typically paused their pace when a generated pipeline appeared in the workspace. 
At these times, some participants used soliloquy, as in saying ``Let me see'', while others kept a focused stare on the workspace. 
These human behaviors suggest that \systemname led participants to engage in deeper, contemplative thought.

The observation suggests that participants needed time to perceive the generated pipelines as they appeared in the workspace.
Such sense-making processes bring new challenges to the creative process:

\myquote{[Using InstructPipe] is a little mentally demanding ... I have to debug ... If it doesn't help (generating an almost 100\% correct pipeline), I have to go through all the nodes ... I don't like debugging.}{P13}

Additionally, we observed that several participants spent more time crafting their prompts than others. 
P15 spent the most time writing the prompt.
The following comments provide insights into how the prompting process caused extra mental workload: 
\myquote{I'm a relatively visual thinker ... Getting the prompt right requires me to think in a way that is a lot more like precise and like getting it figured out without working it out live ... [When writing prompts, ] you’re just putting them (every detail in a whole pipeline) all out [in one short prompt]}{P15}

In addition to the lack of the original visual thinking experience in visual programming, P13 also warned that such simplification of the creative process into prompting experience may sacrifice users' hands-on experiences:
\myquote{I'm very hands-on with techs. I would like to understand what's going on [rather than prompting LLMs to generate everything for me]. I want to like think for myself and then compile all the information myself.}{P13}

\section{Discussion}

\subsection{Human-AI Collaboration in Prototyping Open-ended ML Pipelines}

Our technical evaluation (\cref{sec: tech_results}) shows that \systemname reduces the number of user interactions to 18.9 \% ($\pm 20.3 \%$)
There are two key implications from the results:
\begin{itemize}
    \item \systemname automates \textit{most} pipeline components with a single prompt.
    \item \systemname is \textit{not} able to automate \textit{all} the pipeline creation processes.
\end{itemize}
Such takeaways \textit{differ from early-stage findings} that show LLMs can achieve full automation of ML inference~\cite{suris2023vipergpt, Gupta2023Visual}.
The main reason is that existing work built their ad hoc solutions for target use scenarios, respectively.
In contrast, \systemname covers a larger range of ML models (\cref{sec: nodes}) and aims for an open-ended use case.
Our results show that LLMs (we used GPT-3.5-turbo in the study) still fail to write robust code with prompt engineering techniques.
This aligns with the latest research findings that show that even the latest LLMs still fail to formulate a whole working pipeline~\cite{wen2024programsynthesisbenchmarkvisual, qin2023toolllmfacilitatinglargelanguage}.

While LLMs cannot generate a fully executable pipeline, our work shows that AI can successfully render a certain portion of a pipeline for users.
Both technical and user evaluations highlight the important values here.
We believe such values provide useful takeaways for HCI researchers to explore more human-AI collaboration approaches and designs in visual programming.

\subsection{Three Attributes to Mental Workload}
\label{sec: mental_workload}

Results in \cref{sec: workload} show that \systemname failed to significantly reduce novice users' mental demand.
We summarized its major causes into three aspects.
\subsubsection{\textbf{Instruction}} P15's comment in \cref{sec: feature_request} summarizes the first aspect that causes mental burden.
Although the ``instruction-to-pipeline'' process is fast and seems effortless, the process of framing a prompt is one factor that may overwhelm users, especially those who are more accustomed to visual thinking.
\systemname requires its users to 1) be clear about the problem they want to solve, and 2) be able to clearly articulate the desired pipeline.
Such requirements cause a mental burden to the user~\cite{blackwell1996metacognitive}.
We believe that our results can reinforce the existing knowledge on how non-experts may not prompt LLMs well~\cite{Zamfirescu-Pereira2023Why, nguyen2024how} in the visual programming domain.

\subsubsection{\textbf{Perception}} The integration of LLM support into the visual programming interface enables a ``multimodal programming'' experience~\cite{dietz2023visual}, in which, users can program through both verbal and visual approaches.
However, this flexibility increases perceptual burden as users switch between visual and text-based thinking~\cite{paivio1991dual}.
Interestingly, our results seemingly contradict psychological findings based on the Dual Coding Theory (DCT) that show a combination of verbal and visual information actually helps humans' memory process\footnote{For example, people feel it easier to remember a new word if they learn the word using a vocabulary card with a figure that explains the texts.}~\cite{paivio1969mental, paivio2006dual}.
Therefore, we believe that the mental workload stems not from dislike of multimodal workspaces, but from the lack of a transparent interface that aligns users' mental models with AI reasoning both verbally and visually. 
That being said, a next-generation copilot should visualize a pipeline (\textit{i.e.}, visual info) while the user is prompting the system (\textit{i.e.}, verbal info), constituting and interfacing multimodal processing in humans' brains.

\subsubsection{\textbf{Debugging}}
When a rendered pipeline does not match users' expectations, users need to debug (see P13's comment \cref{sec: feature_request}).
Specifically, users need to \textit{``invest extra effort to review and understand the generated content''}~\cite{zhang2023visar} and then solve the issues caused by the AI assistant.
In essence, debugging is a professional programming skill, which understandably can be mentally overwhelming for beginner-level users.
While \systemname visualizes generated code in the visual programming platform, our results suggest that future systems should provide more guidance for beginners to better proceed with their programming tasks.

\subsection{Instructing LLMs Poses Challenges for Both Novices and, Potentially, Experts}
\label{sec: instructing_challenge}

\begin{figure*}
    \centering
    \begin{subfigure}[b]{0.33\linewidth}
        \includegraphics[width=\linewidth]{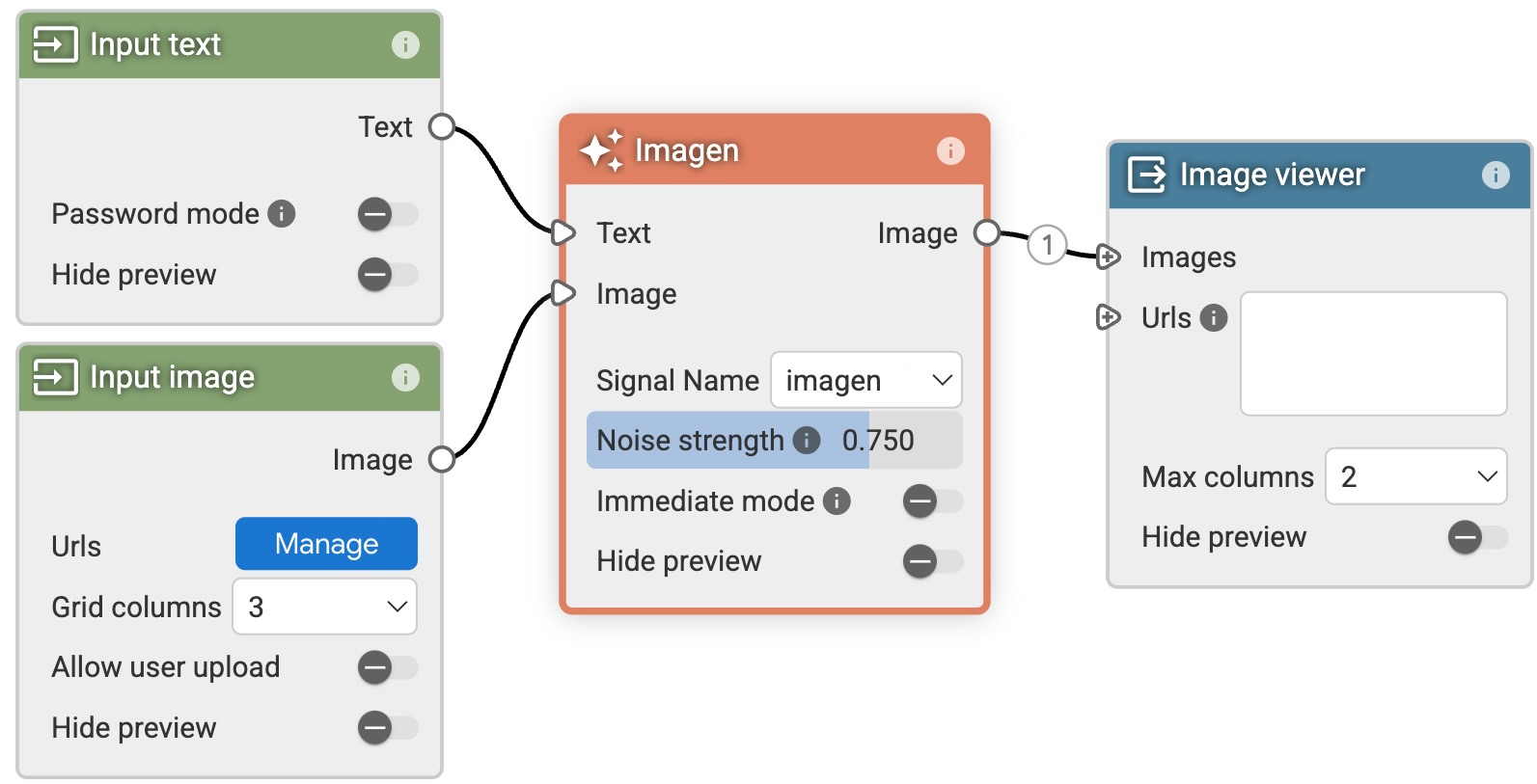}
        \caption{}
        \label{fig: dicussion_short_caption}
    \end{subfigure}
    \hfill
    \vline
    \hfill
    \begin{subfigure}[b]{0.65\linewidth}
        \includegraphics[width=\linewidth]{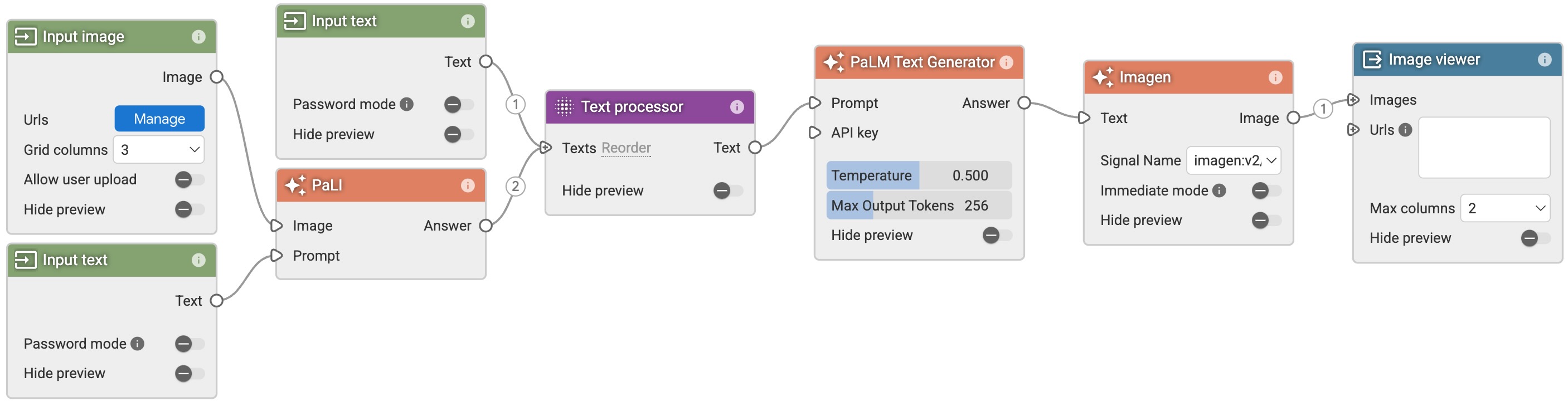}
        \caption{}
        \label{fig: dicussion_long_caption}
    \end{subfigure}
    \centering
    \caption{A comparison of \systemname generated by two instructions: (a) \textit{``Edit an image by updating the image caption''}; (b) \textit{``Caption a tiger image using VQA, modify the character in the caption into a cat using LLM, and finally generate a cat image based on the updated caption''}. See \autoref{fig: showcase_cat} for the complete pipeline.}
    \label{fig: discussion_two_captions}
\end{figure*}

As we discussed above, non-experts found it challenging to instruct LLM.
More interestingly, we found that \textit{even we, the inventors of \systemname, failed to write optimal instructions}.
For instance, the two captions of \autoref{fig: showcase_cat} are \textit{``Describe the image and turn it into a cat image''} and \textit{``Edit an image by updating the image caption''}.
Neither caption explicitly describes the detailed pipeline flow clearly, and therefore, all the six evaluation trials (\cref{sec: tech_evaluation}) were incomplete (see \autoref{fig: dicussion_short_caption} for one example). 
The average ratio of user interactions is $45.8\%$, 
more than twice the average value for our multimodal pipelines ($20.8\%$).
To further understand the cause of the failure, another author improved the instruction into \textit{``Caption a tiger image using VQA, modify the character in the caption into a cat using LLM, and finally generate a cat image based on the updated caption''}.
The resulting pipeline is significantly improved but still not perfect (\autoref{fig: dicussion_long_caption}).
The user only needs to turn ``Imagen'' into another mode so that it also accepts the input ``image'' node.
Revisiting the improved instruction, we instructed \systemname with \textit{``generate a cat image based on THE updated caption''}, which actually missed the input image. 

The important takeaway is while natural languages are proven to be one promising communication media that connects humans and AI systems~\cite{chen2023from, wang2023enabling}, \textit{instructions may not be the best format to facilitate such connection}.
We believe the reason is that instructions are still not intuitive to humans: AI typically requires flawless and unambiguous instructions, while humans tend to express their intentions using ambiguous natural languages in conversations.
We encourage future work to investigate alternative interaction mediums beyond instructions to further enhance user experience in human-AI collaboration.

\section{Limitations and Future Work}
\label{sec: limitations}
\subsection{\reviseSubSectionTitle{Assisting Humans to Prompt AI Copilot in Visual Programming}}

\reviseParagraph{
\systemname introduces a novel user interaction technique for visual programming, along with its set of challenges -- prompting AI is not easy~\cite{Zamfirescu-Pereira2023Why}.
While the latest research has explored prompt writing assistants~\cite{brade2023promptify, promptim, ma2024teach}, creating such assistants in visual programming poses unique challenges, as discussed in~\cref{sec: mental_workload}, and requires further dedicated investigation.
Despite these challenges, the visual programming workspace offers a unique opportunity --  it provides an interactive and visual medium for delivering AI-generated information.
For example, a prompt writing assistant could provide ``a pipeline preview'' in real time via a lightweight LLM.
Visualizing estimated outcomes, such as unexpected pipeline results (as illustrated in~\autoref{fig: dicussion_short_caption}), highlights the potential of these tools to guide users in refining their instructions effectively.
}

\subsection{Node Parameter Tuning}
\systemname focuses on generating the graph structure in the pipeline (\cref{sec: pipeline_representation}), and \systemname is not able to generate node parameters.
The latest research in AI agents shows great potential for distributing a systematic task among multiple LLMs and letting them solve the problem collaboratively~\cite{jin2024llmsllmbasedagentssoftware}.
We encourage future work to extend such distributed AI agent approaches to generate suitable node parameters to further reduce users' workload in tuning them.

\subsection{\reviseSubSectionTitle{A Larger and Dynamic Node Library}}
\systemname is an AI assistant prototype on a small-scale library with 27 nodes.
Similar to other tool-calling LLM systems~\cite{schick2024toolformer, DeLaTorre2024LLMR}, InstructPipe cannot generate any out-of-scope node, and thus, there is a limited scope of pipelines that InstructPipe can generate.
Future work should investigate a scale-up problem by creating an assistant that supports large-scale nodes~\cite{qin2023toolllm}.
What new technical challenge will emerge? 
Will such a large-scale library provide practical human value?
If yes, what are the concrete new user experience it opens up in visual programming?

Additionally, future work should explore a dynamic solution of the node library, in which an AI assistant can define necessary nodes in visual programming on the fly.
HuggingGPT~\cite{shen2024hugginggpt} is a pioneering project that shares a similar vision as this goal, but existing investigations show that the accuracy of such open-ended generation is still unsatisfactory~\cite{patil2023gorilla, qin2023toolllmfacilitatinglargelanguage}.
How can we design an interface to bridge such imperfect AI and human users in visual programming copilot? What will be the interaction paradigm in an interactive system with a dynamic node library?

\subsection{Refining System Component Design}

\systemname provides a system contribution, and we verified the usefulness of \systemname via two evaluations that assess \systemname as a whole system.
One important future direction would be to verify (or even challenge) each technical component of our system\reviseParagraph{, as elaborated below:}

\textbf{Pseudocode}.
We designed the pseudocode order based on how algorithm papers present their algorithm blocks.
Is this design the best approach among all the possible candidates?
If not, how can we further improve the design of pseudocode language?

\reviseParagraph{\textbf{Prompt Design}.
We leveraged the in-context learning capability of LLMs in our prompt design. Prior work shows that few-shot examples cause bias effects in practice~\cite{zhao2021calibrate}, and thus, we encourage future work to mitigate this bias by collecting a large dataset and finetuning LLM on the dataset.}

\reviseParagraph{\textbf{Divide-and-conquer at Scale}.
We adopt the strategy of divide-and-conquer~\cite{smith1985design} with a two-stage LLM pipeline.
Despite its effectiveness with a small node library and simple graphs, its effectiveness is unknown when generating complex graphs.
Exploring agent-based approaches~\cite{huang2023agentcoder, islam2024mapcoder} with Retrieval Augmented Generation (RAG) would be a promising future direction to manage complex graph generation in a divide-and-conquer manner~\cite{smith1985design}.
We encourage future work to contribute high-quality datasets as well as an interactive LLM system with RAG that provides users with better experiences from AI agents.
}

\subsection{\reviseSubSectionTitle{Evaluation Metrics and }Long-term Evaluation}

\reviseParagraph{In the technical evaluation, we assessed the performance of AI assistants based on the number of user interactions.
Existing related metrics, predominantly from the code synthesis literature
~\cite{amini2019mathqa, hendrycks2021measuring}, largely focuses on two categories: correctness-based metrics ~\cite{chen2021evaluating, austin2021program} that rely on test case verification, and similarity-based metrics~\cite{tran2019does}.
\textit{Very little research} falls outside these two categories~\cite{paul2024benchmarks}.
Our metric incorporates human factors by objectively estimating user effort through graph theory, addressing a gap in the visual programming literature where human-centric considerations are crucial.
While our work advances metric development in this domain, further formal research is essential to establish comprehensive standards for visual programming evaluation.}

\reviseParagraph{In the user evaluation,} we conducted a lab study to understand the user experience of \systemname.
As future work, we plan to conduct longer-term studies and gather more realistic insights than those we obtained from the lab study. 
This is critical for us to understand the long-term usefulness of our assistant for beginners, as well as collect feedback to inform our system design.

\subsection{Responsible AI}
\systemname currently cannot detect harmful data or misuse of AI. 
We believe such safety features are crucial, especially in the context of the potential for future dynamic node libraries, which would greatly enhance the generalizability of ML pipeline prototyping capability.
Future work must study effective methods to eliminate potential harmful uses when AI assistants become increasingly open-ended.

\section{Conclusion}

This paper introduces InstructPipe, an AI assistant that empowers users to accelerate their design of ML visual programming pipelines using text instructions.
We design and implement \systemname by decomposing the task into three modules: a node selection module, a code writer, and a code compiler.
Results in our technical and system evaluations suggest that \systemname provides users' satisfactory ``on-boarding'' experience of visual programming systems and allows them to rapidly prototype an idea, potentially with more than 5X fewer interactions.
We further discuss the issues we observed concerning LLMs in visual programming, related to both human factors and technical implementations. 
We hope that InstructPipe will inspire the community to continue work in accelerated human-AI collaboration for increased expressivity and creativity, for machine learning pipelines, and beyond.

\begin{acks}
We would like to thank Gong Xuan, Fengyuan Zhu, Kevin Zhang and Karl Rosenberg for their feedback and discussion on our early-stage prototype, as well as Koji Yatani and Takeo Igarashi for the feedback on the paper draft.
We also thank our reviewers for their insightful feedback.
\end{acks}

\bibliographystyle{ACM-Reference-Format}

\newpage

\clearpage
\appendix
\section*{Appendix}

\section{A Full List of 27 Nodes in InstructPipe}
\label{sec: supp-nodes}
The following content shows 27 nodes \systemname covers in the generation process and their corresponding short description used in the Node Selector (\cref{sec: node_selection}):

\subsection{Input Nodes}
\begin{enumerate}
    \item \textbf{live\_camera}: Capture video stream through your device camera.
    \item \textbf{input\_image}: Select images to use as input to your pipeline. You can also upload your own images.
    \item \textbf{input\_text}: Add text to use as input to your pipeline.
\end{enumerate}
\subsection{Output Nodes}
\begin{enumerate}
    \item \textbf{image\_viewer}: View images.
    \item \textbf{threed\_photo}: Create a 3D photo effect from depthmap tensors.
    \item \textbf{markdown\_viewer}: Render Markdown strings into stylized HTML.
    \item \textbf{html\_viewer}: Show HTML content with styles
\end{enumerate}
\subsection{Processor Nodes}
\begin{enumerate}
    \item \textbf{google\_search}: Use Google to search the web that returns a list of URLs based on a given keyword; usually selected with string\_picker.
    \item \textbf{body\_segmentation}: Segment out people in images; usually selected with mask\_visualizer.
    \item \textbf{tensor\_to\_depthmap}: Display tensor data as a depth map.
    \item \textbf{portrait\_depth}: Generate a 3D depth map for an image; usually selected with tensor\_to\_depthmap, threed\_photo.
    \item \textbf{face\_landmark}: Detect faces in images. Each face contains 468 keypoints; usually selected with landmark\_visualizer, virtual\_sticker.
    \item \textbf{pose\_landmark}: Generate body positional mappings for people detected in images; usually selected with landmark\_visualizer.
    \item \textbf{image\_processor}: Process an image (crop, resize, shear, rotate, change brightness or contrast, add blur or noise).
    \item \textbf{text\_processor}: Reformat and combine multiple text inputs.
    \item \textbf{mask\_visualizer}: Visualize masks.
    \item \textbf{string\_picker}: Select one string from a list of strings; usually used with google\_search.
    \item \textbf{image\_mixer}: Combine images and text into one output image. Requires two image inputs.
    \item \textbf{virtual\_sticker}: Use face landmarks data to overlay virtual stickers on images.
    \item \textbf{palm\_textgen}: Generate Text using a large language model.
    \item \textbf{keywords\_to\_image}: Search for images by keywords.
    \item \textbf{url\_to\_html}: Crawl the website by a given URL.
    \item \textbf{image\_to\_text}: Extract text from an image using OCR service.
    \item \textbf{pali}: Answer questions about an image using a vision-language model.
    \item \textbf{palm\_model}: Generate text using a large language model based on prompt and context.
    \item \textbf{imagen}: Generate an image based on a text prompt.
    \item \textbf{input\_sheet}: Read string data from Google Sheets.
\end{enumerate}

\section{System Implementation}

\subsection{System Prompts Used in LLM Modules}

\begin{figure}
    \centering
    \begin{subfigure}[b]{0.99\linewidth}
        \centering
        \includegraphics[width=\linewidth]{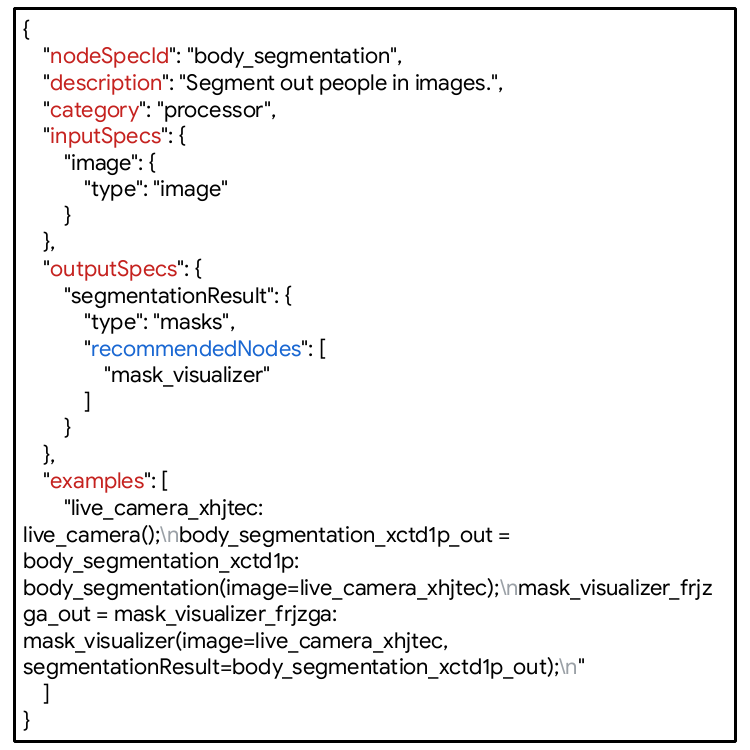}
        \caption{Body segmentation} 
        \label{fig: body}
    \end{subfigure}
    \hfill
    \begin{subfigure}[b]{0.99\linewidth}
        \centering
        \includegraphics[width=\linewidth]{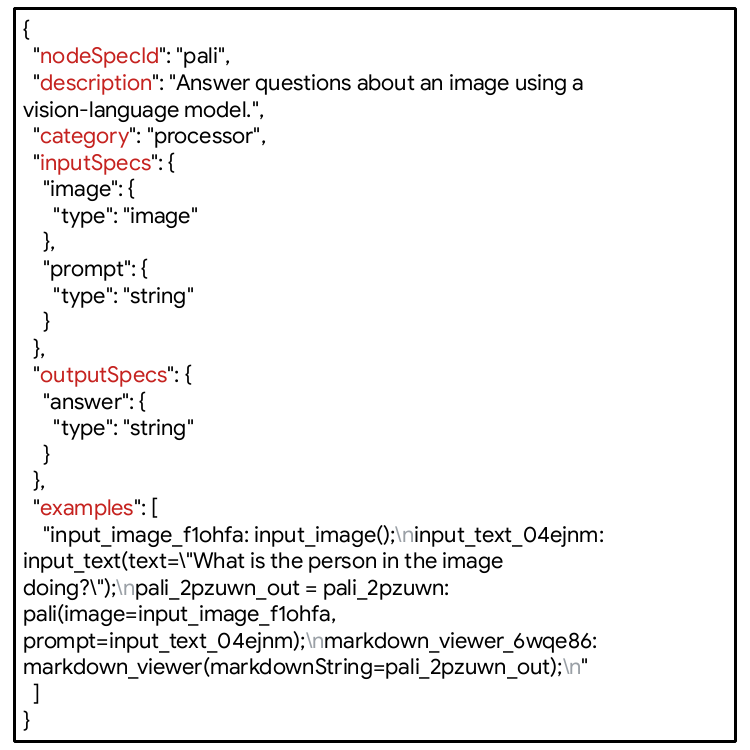}
        \caption{PaLI} 
        \label{fig: pali}
    \end{subfigure}
    \caption{Examples of node configuration used in Code Writer.
    The configuration is structured in a JSON format.
    }
    \label{fig: node_config}
\end{figure}
Here we provide more details about the prompts we utilized in InstructPipe.
The original txt files are also attached in the supplementary zip file.
\subsubsection{Node Selector}
\label{appendix: node_selection}

Please see our supplementary file (node\_select.txt) for the full prompt we use in this stage.

\subsubsection{Code Writer}
\label{appendix: code_writer}
The prompt in Code Writer is dynamic, which is dependent on the nodes chosen in Node Selector.
Therefore, we cannot provide all the possible prompts in the supplementary materials.
Here, we will focus on providing examples of two detailed node configurations utilized in \systemname.
\autoref{fig: code_writer} shows the structure of the prompt utilized in this LLM stage.
\autoref{fig: node_config} provides two examples of node configurations (\ie, ``Body segmentation'' and ``PaLI'') that \systemname may chose into the highlighted line(s).
Each configuration includes keys of ``nodeSpecId'' (\ie, node types), ``description'', ``category'' and ``examples''.
For those nodes that support input and output edges, ``inputSpecs'' and ``outputSpecs'' specify the sockets and their corresponding valid data types.
For example, the output socket name of ``Body segmentation'' is ``segmentationResult'', and its data type is ``masks''.
Some nodes (\eg, ``Body segmentation'') include recommended node(s) (\eg, ``Mask visualizer'' for ``Body segmentation''), and our configuration also contains such information in the dictionary.

\begin{figure*}
    \centering
    \begin{subfigure}[b]{0.4\linewidth}
        \includegraphics[width=\linewidth]{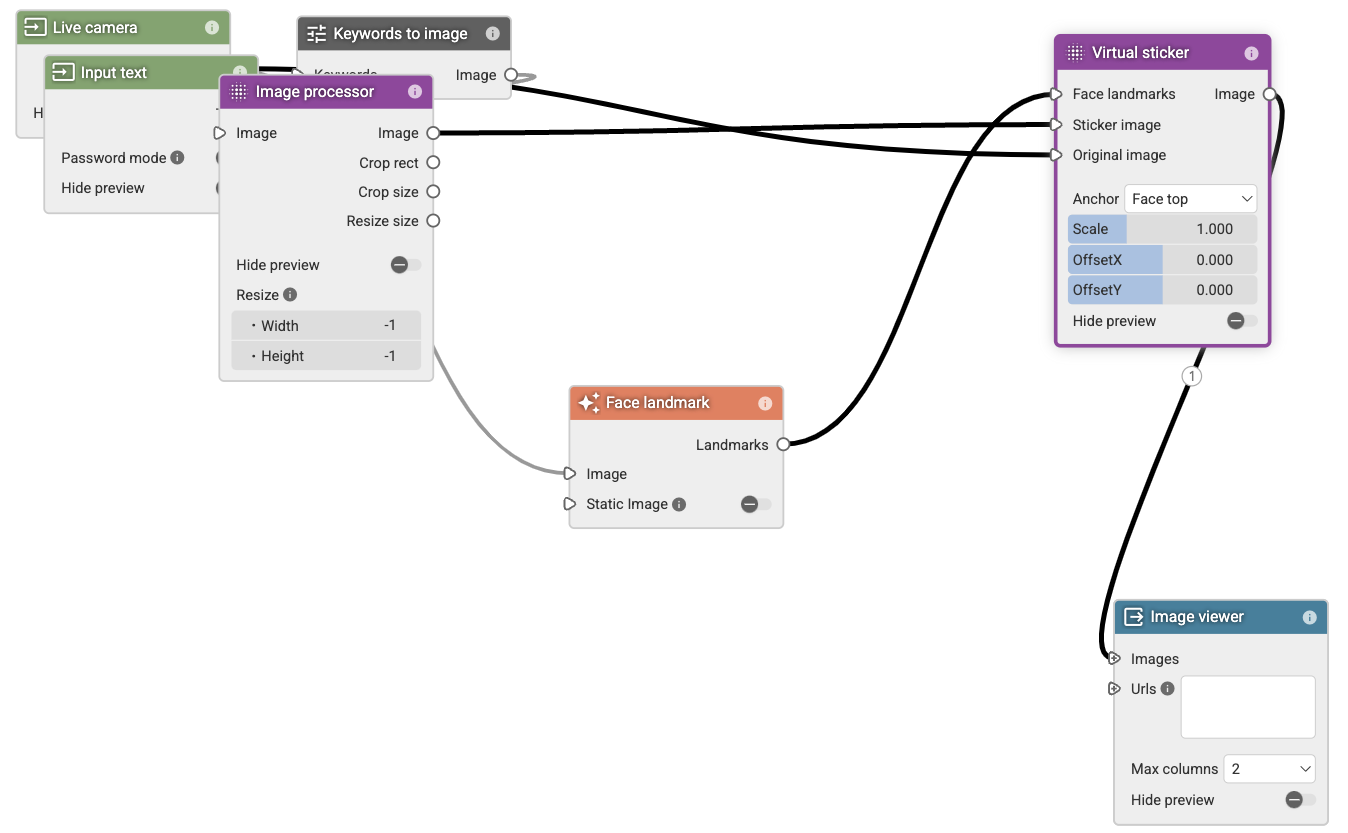}
        \caption{Before layout optimization.}
        \label{fig: smartlayout_before}
    \end{subfigure}
    \hfill
    \hfill
    \begin{subfigure}[b]{0.5\linewidth}
        \includegraphics[width=\linewidth]{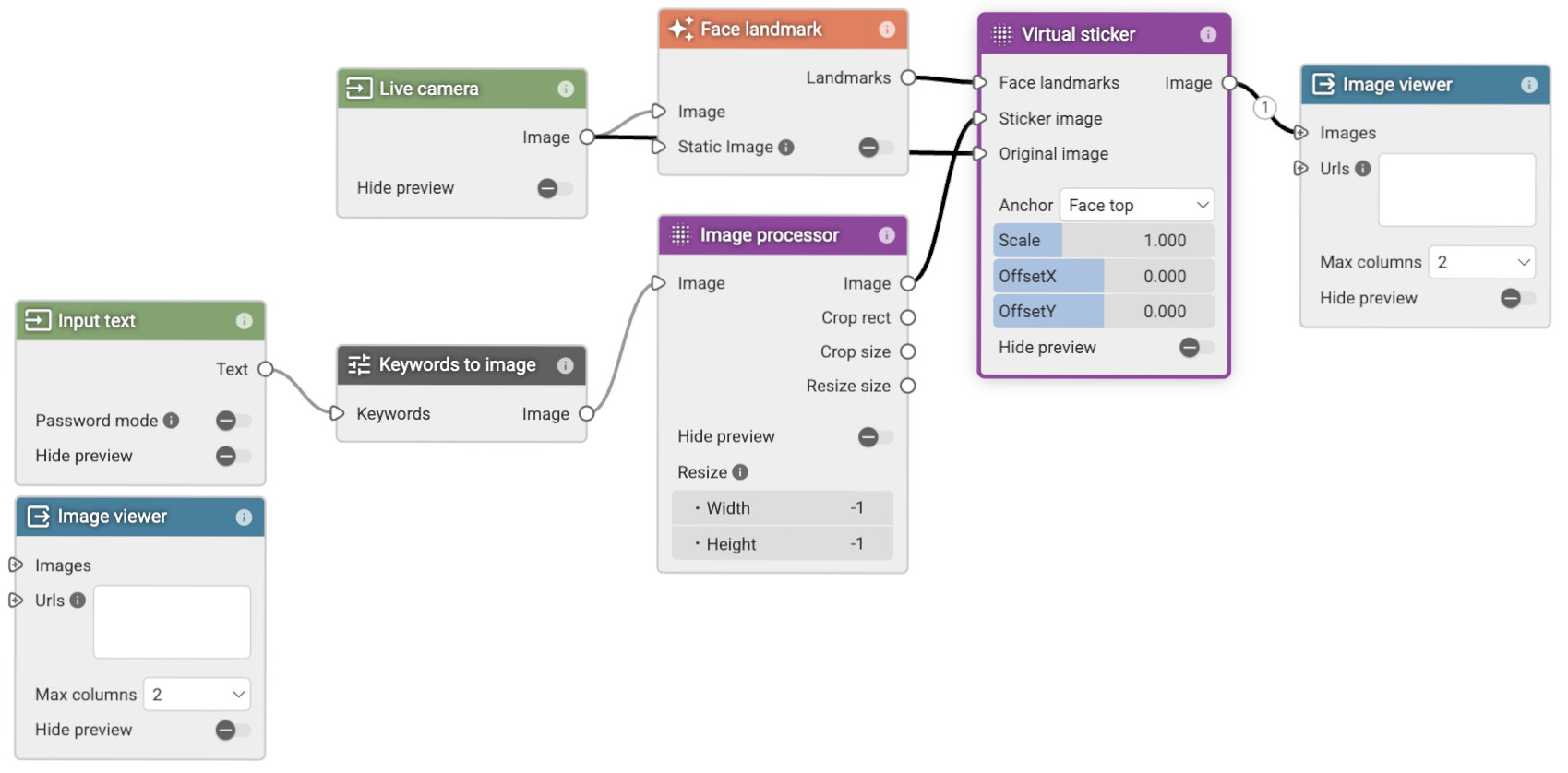}
        \caption{After layout optimization.}
        \label{fig: smartlayout_after}
    \end{subfigure}
    \centering
    \caption{A comparison of the same generated pipeline before and after layout optimization.}
    \label{fig: smartlayout}
\end{figure*}

\subsection{Code Interpreter}
\label{appendix: code_interpreter}
\reviseParagraph{Here, we provide more low-level implementation details on Code Interpreter. 
The Code Interpreter parses generated pseudocode into a visual programming pipeline for visualization at the Visual Blocks workspace.
\autoref{fig:typescript-code} shows the data type definition of graphs, nodes, and edges in the system.
The example JSON file to be parsed into the Typescript interface is available at the Visual Blocks website\footnote{\url{https://visualblocks.withgoogle.com/}. The JSON file is available for data structure exploration by 1) entering an example project and 2) clicking on the ``Export'' button on the top-right corner.}.
Our code defines a visual programming pipeline into an array of serialized nodes, $G(N)$.
When the user adds a new node to a pipeline, it adds a new ``SerializedNode'', containing the edge definition between this new node and other nodes in the current workspace, to the current ``SerialedGraph''.
This mechanism ensures that nodes can be incrementally added in the order they appear in the pseudocode order while maintaining the integrity of the graph by clearly defining dependencies and data flow between nodes.
Algorithm~\ref{alg:code_interpreter} further shows how \systemname parses code and incrementally adds nodes to formulate a final serialized graph.}

\definecolor{LightGray}{gray}{0.9}
\begin{figure}[ht]
\begin{minted}
[
framesep=2mm,
baselinestretch=1,
bgcolor=LightGray,
fontsize=\small,
]
{typescript}
/** A serialized graph.  */
export declare interface SerializedGraph {
  nodes: SerializedNode[];
  
  /** other properties */
}

/** A serialized node.  */
export declare interface SerializedNode {
  /** The id of the node, e.g., pali_1. */
  id: string;

  /** The node spec id, e.g., pali. */
  nodeSpecId: string;
  
  /**
   * Serialized incoming edges that
   * connect to this node.
   */
  incomingEdges?: {
    [inputId: string]: SerializedIncomingEdge[]
  };
  
  /** other properties */
}

/** A serialized incoming edge. */
export declare interface SerializedIncomingEdge {
  /** The id of the source node. */
  sourceNodeId: string;

  /** The id of the output in the source node. */
  outputId: string;
}
\end{minted}
\caption{\reviseParagraph{The definition of a graph, a node and an edge in the system using the Typescript language.
Only the core properties of graphic structure definition are presented in this figure.}}
\label{fig:typescript-code}
\end{figure}

\begin{algorithm}
\caption{\reviseParagraph{Code Interpreter}}
\label{alg:code_interpreter}

\textbf{Input:} $C$: the generated texts (\ie, pseudocode) in the string format. \\
\textbf{Output:} $G(N)$: a visual programming pipeline ($SerializedGraph$) that mainly stores an array of $SeirializedNode$ (\autoref{fig:typescript-code}).\\
\textbf{Variables:} $T$: a dictionary of parsed tokens that contains \textcolor{gred5}{$output\_variable\_id$}, \textcolor{gyellow5}{$node\_id$}, \textcolor{ggreen5}{$node\_type$}, \textcolor{gblue5}{$node\_arguments$}; $e$: the incoming edges of a new node, in the format of $SerializedIncomingEdge$; $n$: a new node in the format of $SerializedNode$.
\BlankLine %
$G: SerializedNode[] = []$ \tcp{Initialize $G$ as an empty array}
$lines = line\_parser(C)$ \tcp{Parse $C$ into lines of code with no pseudocode order changed.}
\BlankLine
\For{$line$ in $lines$}{
    \tcc{
    Example: \\
    $pali\_1\_out = pali\_1: pali(image=input\_image\_1, prompt=input\_text\_1)$ \\
    --> \\
    $\textcolor{gred5}{\text{`}pali\_1\_out\text{'}}$, $\textcolor{gyellow5}{\text{`}pali\_1\text{'}}$, $\textcolor{ggreen5}{\text{`}pali\text{'}}$ and $\textcolor{gblue5}{[\text{`}image=input\_image\_1\text{'}, \text{`}prompt=input\_text\_1\text{'}]}$}
    $T = tokenizer(line)$ \\
    \BlankLine
    \BlankLine
    $ e: incomingEdges = create\_incoming\_edges(T)$
    \tcp{create incoming edges for the new node}
    \BlankLine
    \BlankLine
    $ n: SerializedNode = create\_node(T, e) $ \tcp{create a new SerializedNode with the incomingEdges and the parsed dictionary}
    \BlankLine
    \BlankLine
    $G.add\_serializednode(n)$ \tcp{add the new SerializedNode to the graph}
}
\BlankLine
$G = optim\_layout(G)$ \tcp{Perform the UI layout optimization, as shown in~\autoref{fig: smartlayout}}
\Return $G$

\end{algorithm}

\begin{figure*}[t]
    \centering
    \begin{subfigure}[b]{\linewidth}
        \includegraphics[width=\linewidth]{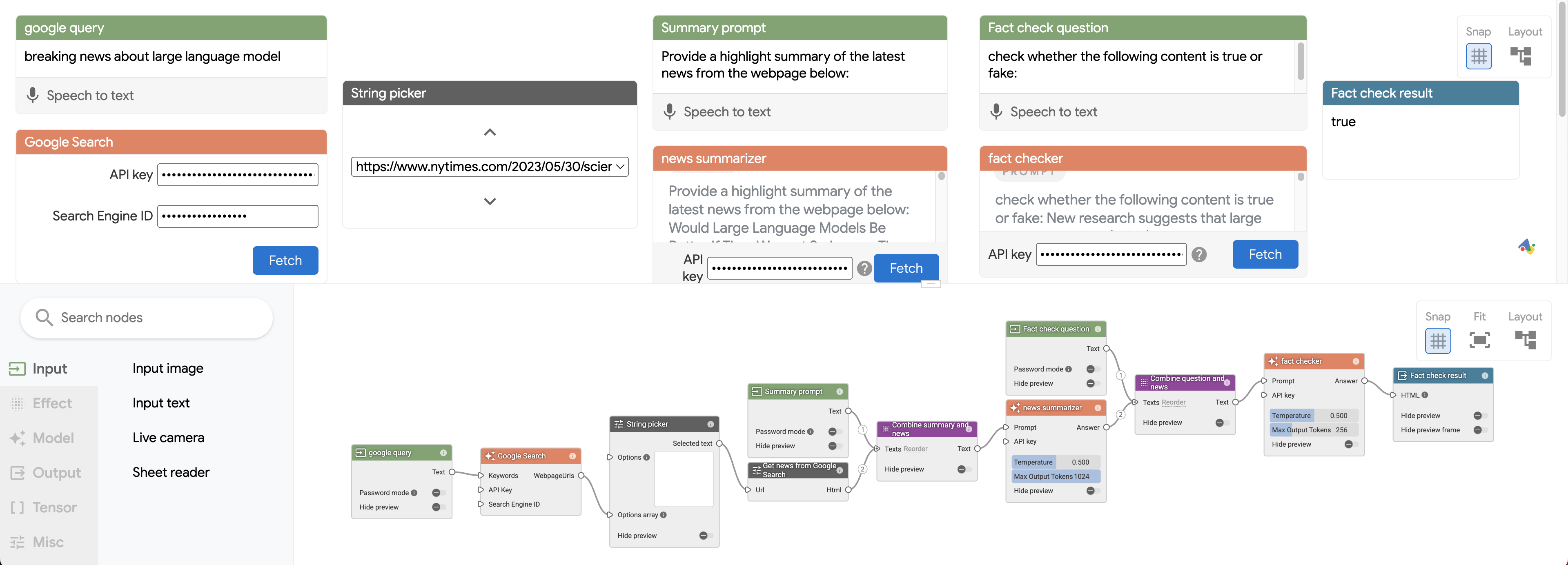}
        \caption{Search news from Google, summarize it, and then conduct a fact check. Input: a keyword for Google Search; Output: a summarization of the news and a fact-check result.}
        \label{fig: showcase_fact_check_news}
    \end{subfigure}
    \begin{subfigure}[b]{\linewidth}
        \includegraphics[width=\linewidth]{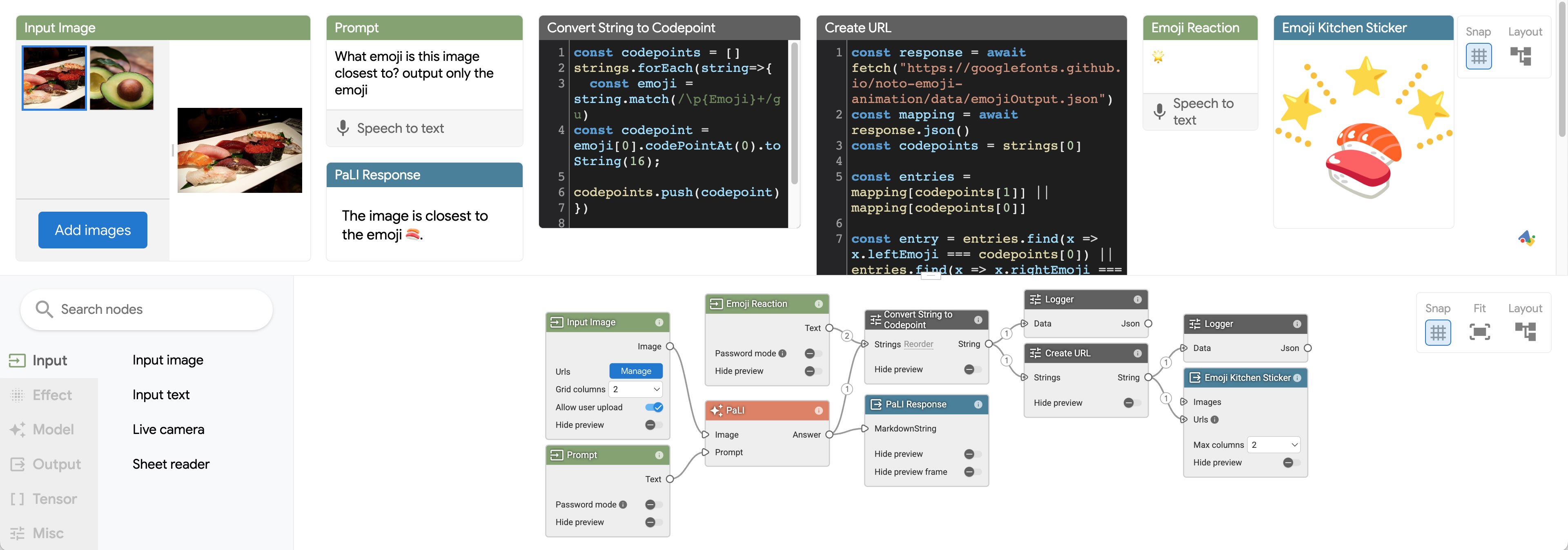}
        \caption{Generating an emoji from a photo. Input: a photo uploaded by the user; Output: an emoji generated from the photo.
        }
        \label{fig: showcase_image_reaction_emoji_kitchen}
    \end{subfigure}
    \begin{subfigure}[b]{\linewidth}
        \includegraphics[width=\linewidth]{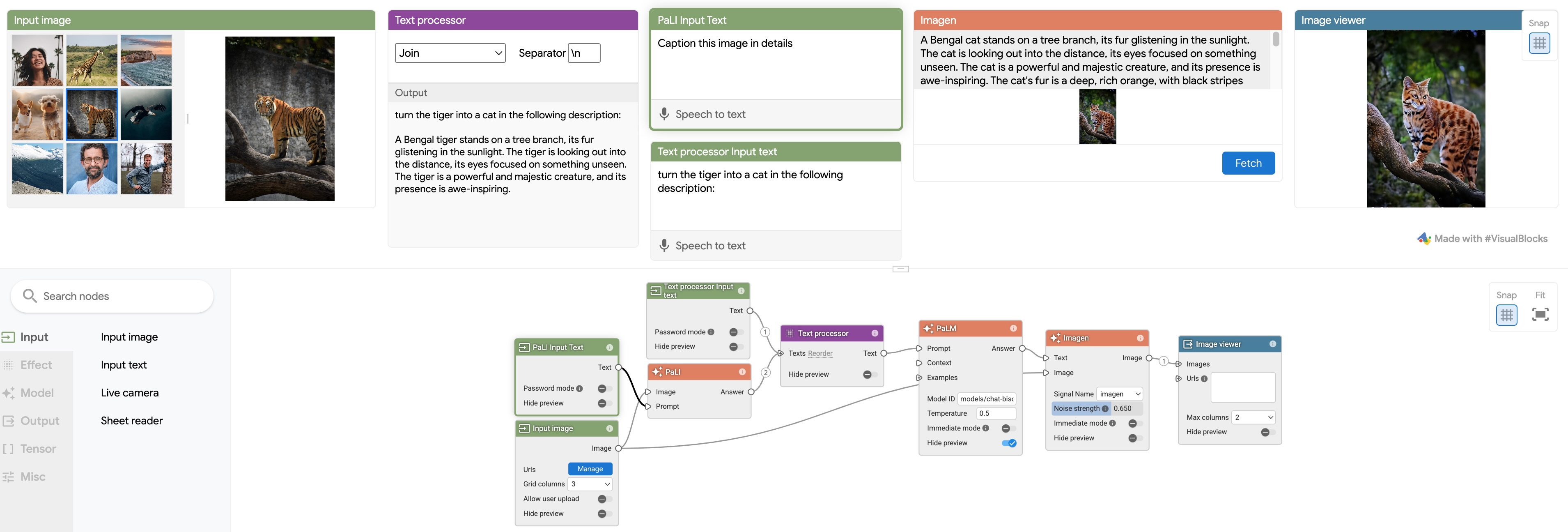}
        \caption{Turning a tiger into a cat. Input: an image of a tiger; Output: an image of a cat.
        }
        \label{fig: showcase_cat}
    \end{subfigure}
    \caption{Example pipelines participants built in the workshops. 
    }
    \label{fig: showcase}
\end{figure*}
\section{Technical Evaluation}

\subsection{Data Post-Processing}
\label{sec: tech_postprocess}

After the workshops, one author carefully examined each collected pipeline and found several critical issues in the raw data:
\begin{itemize}
    \item \textbf{Incomplete pipelines}. There exist pipelines uploaded by the participants that were incomplete.
    \item \textbf{Isolated graphs}. There exist pipelines that include at least one isolated subgraph. 
    The isolated subgraph, as opposed to the main graph, is defined as a graph (or a node) that has no connection to the main graph in the pipeline that provides the main functionality of the pipeline (\eg, the ``Image viewer'' node on the bottom-left corner of \autoref{fig: smartlayout_after}).
    We observed that some participants typically would like to explore the system by working on a separate sub-space. While we acknowledge its usefulness, leaving such ``redundant'' graphs in the raw data for the evaluation would cause issues when we calculate the number of user interactions (\ie, the metric used in the evaluation that will be defined in the next subsection). 
    \item \textbf{Low-quality captions}. While we explicitly required the participants to write descriptive captions, we found some captions written by the participants were either empty or low-quality (\eg, \textit{``newsletter''}, \textit{``image editing''} and \textit{``[participant name]-demo''}).
\end{itemize}

The observation motivated us to post-process the raw data to present more rigorous evaluation results.
We first removed incomplete pipelines and the isolated graphs in each pipeline (if there are any).

To further enhance the annotation quality, two authors individually annotated the caption of each pipeline separately by referring to the original captions and pipelines authored by the participants.
It is important to note that we finished the workshop and the data annotation task \textit{before} we completed the system implementation.
The two authors had no experience using \systemname before completing the annotation.
We believed this process could effectively enhance the quality of the captions while maintaining the fairness of the technical evaluation.

As we clarified in~\cref{sec: data_collection}, the workshop is designed to be an open-ended creation process.
This indicates that the dataset inevitably includes out-of-scope nodes like ``custom scripts'' (in which the participants write code to process the input data and return custom outputs; see~\autoref{fig: showcase_image_reaction_emoji_kitchen} for an example) and ``TFLite model runner'' (which call a custom TensorFlow model with a URL input of the model in the TF-Hub).

We removed the pipelines that contain node(s) out of our focus 27 nodes, and selected all the remaining pipelines as our final evaluation set.
We argue that this post-processing is critical for reporting a fair accuracy value since \systemname can only generate pipelines based on its known node library.
The final 48 pipelines (out of 64 pipelines) are comprised of 23 language pipelines, seven visual pipelines, and 18 multi-modal pipelines.
\autoref{fig: showcase} shows three pipelines created by the participants.
\autoref{fig: showcase_image_reaction_emoji_kitchen} is an example of the pipelines that include out-of-scope nodes, and therefore are not included in the final 48 pipelines.
In the technical evaluation, we ran our generation algorithm on the pipeline captions six times (three times for each caption $\times$ two captions from two authors for each pipeline) and evaluated the generation results using the metric that will be introduced below.

\subsection{Evaluation Metric: The Number of User Interactions}
\label{appendix:metric}

Our definition of the number of user interactions has two important implications.
First, a complete pipeline after user interaction does not need to be the same as the corresponding pipeline in the dataset.
As long as it fulfills the task described in the caption, we consider the pipeline complete. 
Second, our definition does not consider interactions of modifying the node parameters, \eg, typing in a text box or selecting a value in a drop-down box.
We argue that such interactions are highly node-dependent and are hard to quantify objectively.
More importantly, as we explain in~\cref{sec: pipeline_representation}, the generation of node parameters is out of the scope of this work.

In the technical evaluation with various pipelines, it is unfair to report an averaged \textit{absolute} value of user interactions because the complexity of the pipelines varies dramatically.
For instance, the user may need to make three edits based on a generated result to complete a large pipeline that requires 20 edits from scratch.
In another pipeline, the user also needs to do three edits starting from the generated result, but the whole pipeline only takes three edits to finish.
Averaging these \textit{absolute} values does not provide reasonable insights into how accurate the generation is.
Therefore, we reported an averaged \textit{ratio} of user interactions required to complete a pipeline ``from our generated pipeline'' to that ``from scratch'' as our target metric in the technical evaluation.

\section{User Evaluation}

\begin{table*}[h]
\centering
    \caption{Participant demographics for the user study, showing various demographic characteristics and skills relevant to InstructPipe.}
    \label{fig:ppt_demo}
    \includegraphics[width=0.95\linewidth]{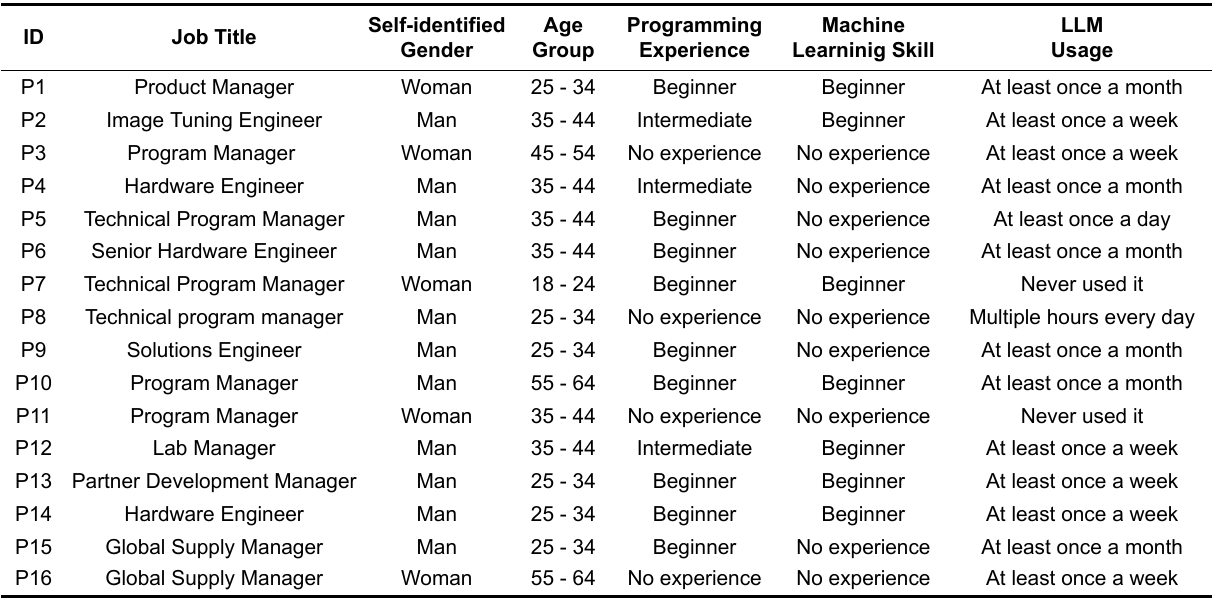}
\end{table*}

\subsection{Semi-structured Interview Script}
\label{sec:supp-interview-script}
~~~~~~[ Introduction ] {\color{gray} ( Start timing! 60 min max. )  }

Hello, my name is X. 

First, I would like to thank you for your participation and completing the consent form. Today, you will be a participant in a user study regarding machine learning and visual programming. Your data will be kept anonymous. Additionally, as a researcher I have no position on this topic and ask that you be as open, honest, and detailed in your answers as possible. Do you have any questions before we begin?

Basically, visual programming borrows the metaphor of block building and allows novice users to develop digital functionalities without writing codes.

[Show Visual Blocks]

Here, each block is called a node, and each node takes in specific inputs, then returns the desired outputs. What you can do is to connect a series of nodes together as a pipeline to achieve a high-level goal.

We are going to walk you through our Visual Blocks system and ask you to actually use Visual Blocks in two conditions to create a few applications.

~~~~~~[ Tutorial ]{\color{gray} ( Start timing! 10 min max. )  }

Before we get started, let us do a tutorial of our system.

~~~~~~[ Study and TLX ]{\color{gray} ( Start timing! about 30 min )  }

[Leverage the counter-balanced sheet and give user a task]

[Think aloud. Have a short discussion with the user. What’s the user’s plan to achieve this given functionality?]

~~~~~~[ Interview ]{\color{gray} ( Start timing! about 15 min )  }

1. What’s your impression of Visual Blocks / InstructPipe [counterbalanced]? Do you need many edits / operations to make it work?

2. Are there any pipelines you come up with in work scenarios / casual scenarios?

3. What works with InstructPipe? In what specific scenarios will InstructPipe be very helpful?

4. What does not work with InstructPipe? Would you give me an example?

5. Do you have any suggestions to improve the design of both systems?

6. Which kinds of technologies would be interesting to add?

7. What applications do you want to start with InstructPipe? And what applications do you want start without it?

That’s all for our user study. Thank you for your participation and we will compensate for your time.

\subsection{User Study Pipelines}
\label{appendix: user-study-pipeline}
\begin{figure*}[h]
    \centering
    \includegraphics[width=\linewidth]{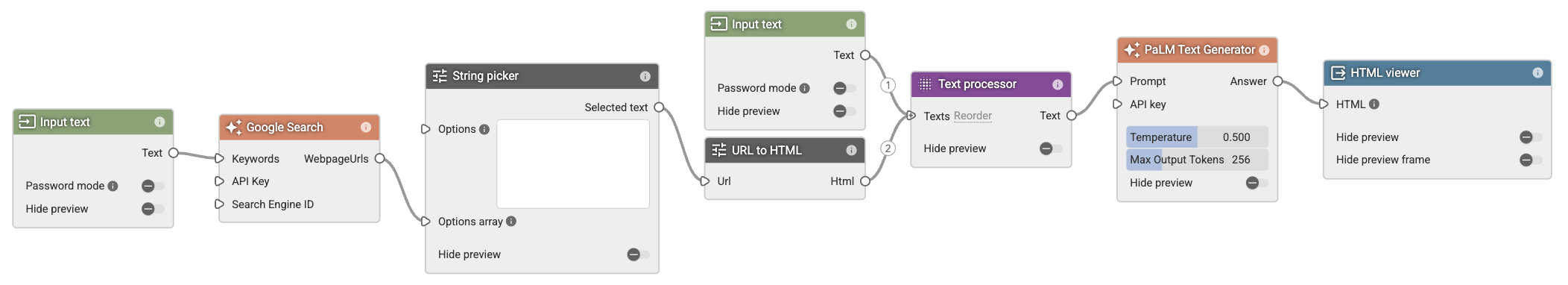}
    \caption{Text-based pipeline. 
    The ``String picker'' node provides users a drop-down menus to select one URL from a list of URLs returned by ``Google Search''.
    ``PaLM Text Generator'' is an LLM used to summarize the full HTML page.
    } 
    \label{fig: text}
\end{figure*}

\begin{figure}
    \centering
    \includegraphics[width=0.7\linewidth]{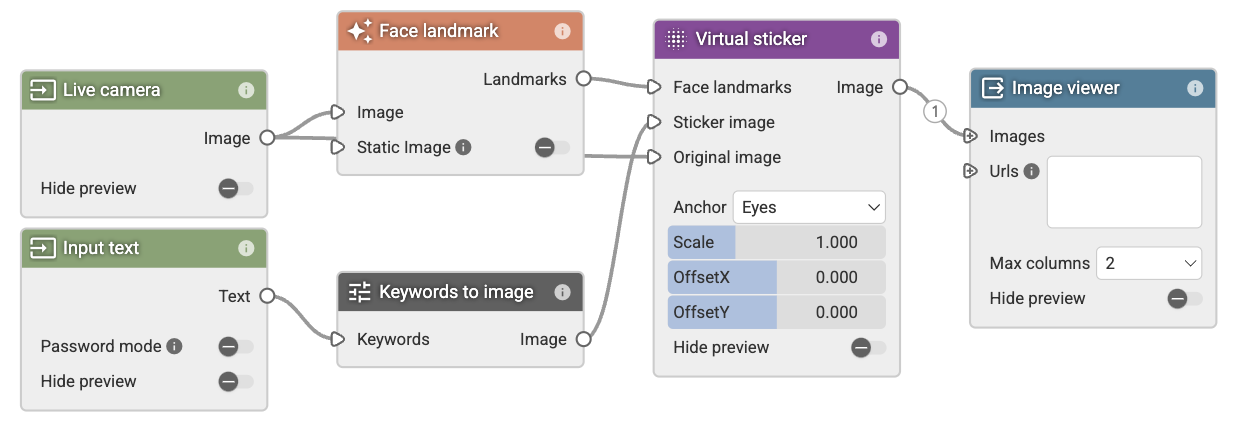}
    \caption{Real-time multimodal pipeline. The ``Keyword to image'' node is used to search a sunglasses image, and the ``Virtual sticker'' node anchors the sunglasses onto the user' face.}
    \label{fig: multimodal}
\end{figure}

\autoref{fig: text} and \autoref{fig: multimodal} visualize two pipelines we required the participants to complete in our user study.
\autoref{fig: multimodal} is a multimodal pipeline that allows participants to interact with AR effects in real time.
Our technical evaluation shows that \systemname can generate this pipeline accurately: the averaged ratio of human interactions = 5.2\%.
\autoref{fig: multimodal} is a text-based pipeline that provides participants with a summary of the news searched from Google.
The technical evaluation reveals that \systemname cannot generate this pipeline accurately without further human interaction, and the average ratio of additional human interactions is 27.8\%.
While the generated diagram (with error) is not deterministic, we observed that InstructPipe commonly generates the pipeline in Figure 14 without ``URL to HTML'' or ``PaLM Text Generator'' nodes.
The error implies that the LLM may misinterpret 1) the data from the ``selected text'' port of the ``String picker'' node is the texts on the web instead of the web URL and 2) that ``Text processor'' has the LLM capability to process the texts instead of simply combining two texts.

Note that even though \systemname may be able to complete the pipeline structure in \autoref{fig: multimodal} from users' instruction, we observed that participants still need to fine-tune their keywords to get an ideal pair of sunglasses.
Additionally, the default anchor value is ``Face top'', so participants need to use the drop-down menu on the ``Virtual sticker'' node to change the value to ``Eyes''. 
This further motivates us to use the metric of ``Time'' in addition to the number of user interactions in our study.
Our demo video also covers the workflows of these two pipelines.

\subsection{Assistant Provided to the Participants in the User Evaluation}
\label{appendix: user_study_assistant}
In the user evaluation, our goal is to make the interface condition (either InstructPipe or Visual Blocks) as the only independent variable that changes our dependent variables (\cref{sec: metric}).
Similar to user evaluations of other early-stage HCI research, we had to improvise for lacking system affordances. As an example, we would include help menus and error recovery models in the future versions of our system, but at this early stage, we relied on in-person help to nudge and assist our user study participants.
We took actions (\ie, assistants) in the user evaluation to ensure the study is under an appropriate amount of control as well as maintain the fairness of our study.

Here, we elaborate on two examples of assistants we provided in the user study. 

In the InstructPipe condition, one participant started their ``instructions'' by dragging a text box into the visual programming workspace and began typing.
When noticing this issue, we kindly asked the participant whether s/he wanted to write instructions or build a pipeline from scratch.
S/he then noticed this issue and clicked on the ``InstructPipe'' button to write prompts.
Note that we explicitly taught every participant how to use InstructPipe and asked participants themselves to go through the instruction processes in the training task (\autoref{fig:user_evaluation_flow}).

In the Visual Blocks condition, one participant first dragged a ``Virtual sticker'' into the workspace when s/he wanted to build the multimodal pipeline as required (\autoref{fig: multimodal}).
After a while, s/he asked us for the meaning of ``landmarks'' on the first input port of the ``Virtual sticker'' node (\autoref{fig: multimodal}).
We then answered this question and provided a hint on the ``Face landmark'' node (\autoref{fig: multimodal}) that could produce the ``Face landmarks'' required by the ``Virtual sticker''.
While we had explained all the nodes that the participants need to use in the study in the training task (\autoref{fig:user_evaluation_flow}), we consider such technical questions reasonable because all of our participants are non-experts.
Programming itself is a difficult skill, and it is quite common that people may forget some of the knowledge that they have just learned.
Instead of being silent and keeping the participants stuck on a technical issue, we believed offering technical help was an important action we must take to ensure the data quality we collected in the study.

These anecdotes in the user evaluation reveal several limitations of the visual programming system: some designs may not be very intuitive to non-experts.
Since the goal of our user evaluation is understanding the benefits of \systemname compared to Visual Blocks (without AI assistants), we made our best efforts to take action to prevent the effects caused by other factors from influencing our data.
Meanwhile, we also encourage future work to further explore the system design so that future users can more easily use our assistant in visual programming.

\begin{table*}[]
\caption{The counterbalance sheet of the user evaluation.
Each cell is in the format of ``Interface / Pipeline''. 
``Instruct'' and ``VB'' mean the ``InstructPipe'' and ``Visual Blocks'' conditions, respectively.
``Search'' and ``Tryon'' represent the ``text-based pipeline'' (\autoref{fig: text}) and the ``real-time multimodal pipeline'' (\autoref{fig: multimodal}), respectively.
}
\label{tab: counterblance}
\begin{tabular}{lllll}
\hline
ID  & \multicolumn{1}{c}{Step 1} & \multicolumn{1}{c}{Step 2} & \multicolumn{1}{c}{Step 3} & \multicolumn{1}{c}{Step 4} \\ \hline
P1  & Instruct / Tryon           & Instruct / Search          & VB / Tryon                 & VB / Search                \\
P2  & VB / Tryon                 & VB / Search                & Instruct / Tryon           & Instruct / Search          \\
P3  & Instruct / Search          & Instruct / Tryon           & VB / Search                & VB / Tryon                 \\
P4  & VB / Search                & VB / Tryon                 & Instruct / Search          & Instruct / Tryon           \\
P5  & VB / Tryon                 & VB / Search                & Instruct / Tryon           & Instruct / Search          \\
P6  & Instruct / Tryon           & Instruct / Search          & VB / Tryon                 & VB / Search                \\
P7  & VB / Search                & VB / Tryon                 & Instruct / Search          & Instruct / Tryon           \\
P8  & Instruct / Search          & Instruct / Tryon           & VB / Search                & VB / Tryon                 \\
P9  & Instruct / Tryon           & Instruct / Search          & VB / Tryon                 & VB / Search                \\
P10 & VB / Tryon                 & VB / Search                & Instruct / Tryon           & Instruct / Search          \\
P11 & Instruct / Search          & Instruct / Tryon           & VB / Search                & VB / Tryon                 \\
P12 & VB / Search                & VB / Tryon                 & Instruct / Search          & Instruct / Tryon           \\
P13 & VB / Tryon                 & VB / Search                & Instruct / Tryon           & Instruct / Search          \\
P14 & Instruct / Tryon           & Instruct / Search          & VB / Tryon                 & VB / Search                \\
P15 & VB / Search                & VB / Tryon                 & Instruct / Search          & Instruct / Tryon           \\
P16 & Instruct / Search          & Instruct / Tryon           & VB / Search                & VB / Tryon                 \\ \hline
\end{tabular}
\end{table*}

\subsection{Counter-Balancing and The Replication Number}
\autoref{tab: counterblance} presents how we perform counterbalance in the user evaluation.
We counterbalanced both the interface factor (``\systemname'' and ``Visual Blocks'') and the pipeline factors to reduce the learning effects. 
We then replicated the order four times so that we collected multiple data from different participants in each unique study order.
This helps strengthen the power of the data we collected in the study.
Note that, in the group of P5 - P8, we flipped the orders within P5 and P6 as well as P7 and P8, but this does not cause a difference in the counterbalance process.

\balance

\end{document}